\documentclass[]{aa}  

\usepackage{txfonts}
\usepackage{mathtools}
\usepackage[usenames]{color}
\usepackage[dvipsnames]{xcolor}

\usepackage{subfig}
\usepackage{booktabs,makecell}
\usepackage{siunitx}

\usepackage{enumitem}
\usepackage{graphicx}
\usepackage{hyperref}
\usepackage{threeparttable}
\usepackage{float}
\usepackage{lscape}
\usepackage{xcolor}
\usepackage{amsmath}
\usepackage{comment}

\definecolor{green}{rgb}{0.,1.,0.}
\definecolor{greenMM}{rgb}{0.,0.8,0.7}
\definecolor{orange}{rgb}{1,0.5,0}

\newcommand\F[1]{{\color{Green} #1}} 
\newcommand\JD[1]{{\color{orange} #1}} 
\newcommand\PC[1]{{\color{red} #1}} 

\begin{document}

   \title{Locating dust and molecules in the inner circumstellar environment of R~Sculptoris with MATISSE}

  
  \titlerunning{Locating dust and molecules around R~Sculptoris with MATISSE}

   \author{
         J. Drevon\inst{\ref{OCA}, \ref{ESOchile}} \fnmsep\thanks{Corresponding author: \email{julien.drevon@oca.eu}}
          \and
          F. Millour\inst{\ref{OCA}}\fnmsep\thanks{Corresponding author: \email{fmillour@oca.eu}}
          \and
          P. Cruzal\`ebes\inst{\ref{OCA}}
          \and
          C. Paladini\inst{\ref{ESOchile}}
          \and
          J. Hron\inst{\ref{vienne}}
          \and
          A. Meilland\inst{\ref{OCA}}
          \and
          F. Allouche\inst{\ref{OCA}}
          \and
          K.-H. Hofmann\inst{\ref{MPIFR}}
          \and
          S. Lagarde\inst{\ref{OCA}}
          \and
          B. Lopez\inst{\ref{OCA}}
          \and
          A. Matter\inst{\ref{OCA}}
          \and
          R. Petrov\inst{\ref{OCA}}
          \and
          S. Robbe-Dubois\inst{\ref{OCA}}
          \and
          D. Schertl\inst{\ref{MPIFR}}
          \and
          P. Scicluna\inst{\ref{ESOchile}}
          \and
          M.~Wittkowski \inst{\ref{ESOgarching}}
          \and 
          G. Zins\inst{\ref{ESOchile}}\and\\
P.~Ábrahám \inst{\ref{konkoly}}
\and P.~Antonelli \inst{\ref{OCA}}
\and U.~Beckmann \inst{\ref{MPIFR}}
\and P.~Berio \inst{\ref{OCA}}
\and F.~Bettonvil \inst{\ref{SRON}}
\and A. Glindemann \inst{\ref{ESOgarching}}
\and U.~Graser \inst{\ref{MPIA}}
\and M.~Heininger \inst{\ref{MPIFR}}
\and Th.~Henning \inst{\ref{MPIA}}
\and J. W. Isbell \inst{\ref{MPIA}}
\and W. Jaffe \inst{\ref{leiden}}
\and L.~Labadie \inst{\ref{cologne}}
\and C.~Leinert \inst{\ref{MPIA}}
\and M.~Lehmitz \inst{\ref{MPIA}}
\and S. Morel \inst{\ref{OCA}}
\and K. Meisenheimer \inst{\ref{MPIA}}
\and A.~Soulain \inst{\ref{OCA}}\fnmsep\thanks{Now at the Sydney Institute for Astronomy (SIfA), School of Physics, The University of Sydney, NSW 2006, Australia}
\and J.~Varga \inst{\ref{leiden},\ref{konkoly}}
\and G. Weigelt \inst{\ref{MPIFR}}
\and J.~Woillez \inst{\ref{ESOgarching}}
\and J.-C.~Augereau \inst{\ref{IPAG}}
\and R.~van~Boekel \inst{\ref{MPIA}}
\and L.~Burtscher \inst{\ref{leiden}}
\and W.C. Danchi \inst{\ref{Goddard}}
\and C. Dominik \inst{\ref{amsterdam}}
\and V. G\'amez Rosas \inst{\ref{leiden}}
\and V. Hocd\'e \inst{\ref{poland}}
\and M.R.~Hogerheijde \inst{\ref{leiden},\ref{amsterdam}}
\and L.~Klarmann \inst{\ref{MPIA}}
\and E.~Kokoulina \inst{\ref{OCA}}
\and J.~Leftley\inst{\ref{OCA}}
\and P.~Stee \inst{\ref{OCA}}
\and F.~Vakili \inst{\ref{OCA}}
\and R.~Waters \inst{\ref{radboud},\ref{SRON}}
\and S. Wolf \inst{\ref{Kiel}}
\and G.~Yoffe \inst{\ref{MPIA}}
}

\institute{Universit\'e Côte d'Azur, Observatoire de la C\^ote d'Azur, CNRS, Laboratoire Lagrange, France\label{OCA}
\and European Southern Observatory, Alonso de Córdova, 3107 Vitacura, Santiago, Chile\label{ESOchile}
\and Department of Astrophysics, University of Vienna, T\"urkenschanzstrasse 17\label{vienne}
\and Max-Planck-Institut f\"ur Radioastronomie, Auf dem H\"ugel 69, D-53121 Bonn, Germany\label{MPIFR}
\and European Southern Observatory, Karl-Schwarzschild-Str. 2, 85748 Garching, Germany\label{ESOgarching}
\and Konkoly Observatory, Research Centre for Astronomy and Earth Sciences, E\"otv\"os Lor\'and Research Network (ELKH), Konkoly-Thege Mikl\'os \'ut 15-17, H-1121 Budapest, Hungary\label{konkoly}
\and Max Planck Institute for Astronomy, K\"onigstuhl 17, D-69117 Heidelberg, Germany\label{MPIA}
\and Leiden Observatory, Leiden University, Niels Bohrweg 2, NL-2333 CA Leiden, The Netherlands\label{leiden}
\and I. Physikalisches Institut, Universit\"at zu K\"oln, Z\"ulpicher Str. 77, 50937, K\"oln, Germany\label{cologne}
\and Univ. Grenoble Alpes, CNRS, IPAG, 38000, Grenoble, France\label{IPAG}
\and NASA Goddard Space Flight Center, Astrophysics Division, Greenbelt, MD 20771, USA\label{Goddard}
\and Anton Pannekoek Institute for Astronomy, University of Amsterdam, Science Park 904, 1090 GE Amsterdam, The Netherlands\label{amsterdam}
\and Nicolaus Copernicus Astronomical Center, Polish Academy of Sciences, Bartycka 18, 00-716 Warszawa, Poland\label{poland}
\and Department of Astrophysics/IMAPP, Radboud University, P.O. Box 9010, 6500 GL Nijmegen, The Netherlands\label{radboud}
\and SRON Netherlands Institute for Space Research Sorbonnelaan 2, 3584 CA Utrecht, The Netherlands\label{SRON}
\and Institut f\"ur Theoretische Physik und Astrophysik, Christian-Albrechts-Universit\"at zu Kiel, Leibnizstra{\ss}e 15, 24118, Kiel, Germany\label{Kiel}
}

   \authorrunning{Drevon et al.}

   \date{Received 22 June 2021 / Accepted 10 June 2022}

 
  \abstract
   {Asymptotic giant branch (AGB) stars are one of the main sources of dust production in the Galaxy. However, it is not yet clear what this process looks like and where the dust happens to be condensing in the circumstellar environment.}
   {By characterizing the location of the dust and the molecules in the close environment of an AGB star, we aim to achieve a better understanding the history of the dust formation process.}
   {We observed the carbon star \object{R Scl} with the thermal-infrared VLTI/MATISSE instrument in $L$- and $N$-bands. The high angular resolution of the VLTI observations (as small as 4.4\,mas in the $L$-band and 15\,mas in the $N$-band with ATs), combined with a large uv-plane coverage allowed us to use image reconstruction methods. To constrain the dust and molecules' location, we used two different methods: one using MIRA image reconstruction algorithm and the second using the 1D code \texttt{RHAPSODY}.} 
   {We found evidence of C$_2$H$_2$ and HCN molecules between 1 and 3.4\,$R_\star$ which is much closer to the star than the location of the dust (between 3.8 and 17.0 $R_\star$). We also estimated a mass-loss rate of $ 1.2\pm 0.4 \times 10^{-6} M_{\odot}~\rm{yr}^{-1}$. In the meantime, we confirmed the previously published characteristics of a thin dust shell, composed of amorphous carbon (amC) and silicon carbide (SiC). However, no clear SiC feature has been detected in the MATISSE visibilities. This might be caused by molecular absorption that can affect the shape of the SiC band at 11.3~$\mu$m.}
   {The appearance of the molecular shells is in good agreement with predictions from dynamical atmosphere models. For the first time, we co-located dust and molecules in the environment of an AGB star. We confirm that the molecules are located closer to the star than the dust. The \texttt{MIRA} images unveil the presence of a clumpy environment in the fuzzy emission region beyond 4.0\,$R_\star$. Furthermore, with the available dynamic range and angular resolution, we did not detect the presence of a binary companion. To solve this problem, additional observations combining MATISSE and SAM-VISIR instrument should enable this detection in future studies.}


   \maketitle
%








\section{Introduction}


\object{R Scl} is a bright infrared source known to be a semi-regular pulsating C-rich asymptotic giant branch (AGB) star (Srb) with a pulsation period of about 370 days \citep{samus2009} and a mass-loss rate (published in the literature) ranging from $2\times10^{-10}\,M_{\sun}~\rm{yr}^{-1}$ \citep{Brunner2018} up to $1.6\times10^{-6}\,M_{\sun}~\rm{yr}^{-1}$ \citep{deBeck2010}. Carbon stars are spectrally classified in R or N-type stars supplementing the classical hotter O, B, A, F, G, K, and M spectral types. While the N-type stars are colder than R stars, they are marked with a number identifying the effective temperature range of the star's location. \citet{Barnbaum1996} classified R~Scl as a C-N5-type star in the revised MK classification of \citet{keenan1993}, with C/O$\approx1.4$ \citep{hron1998}. Its distance is estimated to be $360\pm50$\,pc by \citet{Maercker2018}, and measured as 440$\pm$30\,pc (DR2) and $393\pm12$\,pc (DR3) by \citet{Gaia2018}. These two Gaia distances are marginally compatible within 1.5~$\sigma$ with the values from \citet{Maercker2018}  and with each other; therefore, we retained the conservative value of \citet{Maercker2018} ($360\pm50$\,pc) for the purposes of this work.

R~Scl is one of just a few detached shell objects \citep{olofsson1996} and ALMA observations revealed a spiral structure and the presence of clumpy and non-centrosymmetric structures inside the shell. \citet{Maercker2012} suggested it could be explained by the presence of a so far unseen companion of $0.25\,M_{\odot}$ hidden at $60$\,au (~34\,R$_\star$) in the dusty shell. 
VLTI/PIONIER images do not show that companion, but instead show an extended photosphere with a dominant spot that is very likely to be of a convective origin \citep{wittkowski2017}. Polarimetric observations confirm the clumpy structure of the dust forming region \citep{yudin2002}. However, such clumps were not detected in the MIDI observations of \cite{sacuto2011}. Such a non-detection was very likely due to the limitations in terms of uv-coverage and visibility sampling of MIDI, which is not optimized for the detection of large-scale asymmetric structures \citep{paladini2012, paladini2017}.


This paper is organized as follows. In Section~\ref{sec:data}, we present new VLTI/MATISSE observations in $L$- and $N$-bands of the AGB star R~Scl, as well as the data reduction processes. In Section~\ref{sec:sed}, we present the photometric data of this star and we derived a hydrostatic model to describe the central source. The resulting best-fit model is used as input for the DUSTY modeling used to interpret both the photometry and the interferometric MATISSE data. In Sections~\ref{sec:model} and~\ref{sec:discussion}, we present the very first chromatic images in the $L$ (3--4\,$\mu$m) and $N$ (8--13\,$\mu$m) mid-infrared (mid-IR) bands of the star's photosphere and its immediate vicinity. In addition, to locate the molecules and the dust in the circumstellar shell, we modeled the measured visibility data with a combination of ring-shaped layers within the close environment of the star and discuss our results. Finally, in Section~\ref{sec:conclusion}, we provide our conclusions.



%



\section{Observations and data processing} \label{sec:data}

 The Multi Aperture Interferometric SpectroScopic Experiment (MATISSE) is a  four-telescope interferometric beam combiner that disperses the coherent light from four telescopes of the Very Large Telescope Interferometer (VLTI) in the $L$ (3--4\,$\mu$m), $M$ (4.5--5\,$\mu$m), and $N$ (8--13\,$\mu$m) mid-infrared bands \citep{Lopez2021}.
It can combine either the four 1.8\,m telescopes (auxiliary telescopes, AT), or the four 8\,m telescopes (unit telescopes, UT). MATISSE delivers coherent fluxes, closures phases, differential phases, and photometry which can be used with the coherent flux to derive the relevant visibilities.

The four ATs can be moved along rails on the VLTI site, allowing for the baselines (separation vector between two telescopes with coordinates (u,v)) to more densely populate the (u,v)-plane than when only 4 fixed telescopes are used. This makes MATISSE an imaging instrument that offers an exquisite spatial resolution: reaching, in the $L$-band, up to $\lambda_L=3\,\mu$m, $\lambda_L/B = 4.4$\,mas, and in the $N$-band up to  $\lambda_N=10\,\mu$m, $\lambda_N/B = 15$\,mas for a baseline of $B=140$\,m. The reliability of the image reconstruction process, the signal-to-noise-ratio (S/N), and the dynamic range of the image depend on the completeness of the (u,v)-plane coverage.





The observations of R~Scl were conducted using the ATs of the VLTI during the commissioning run of MATISSE in December 2018. 
Table~\ref{tab:log} presents the logbook of the observations with the used AT configurations. Since on December 5 and 6, the observing nights were of poor quality characterized by technical glitches, only a fraction of the data from these two nights are used here.


\begin{figure}[htbp]
\begin{center}
\includegraphics[angle=-0,width=0.45\textwidth]{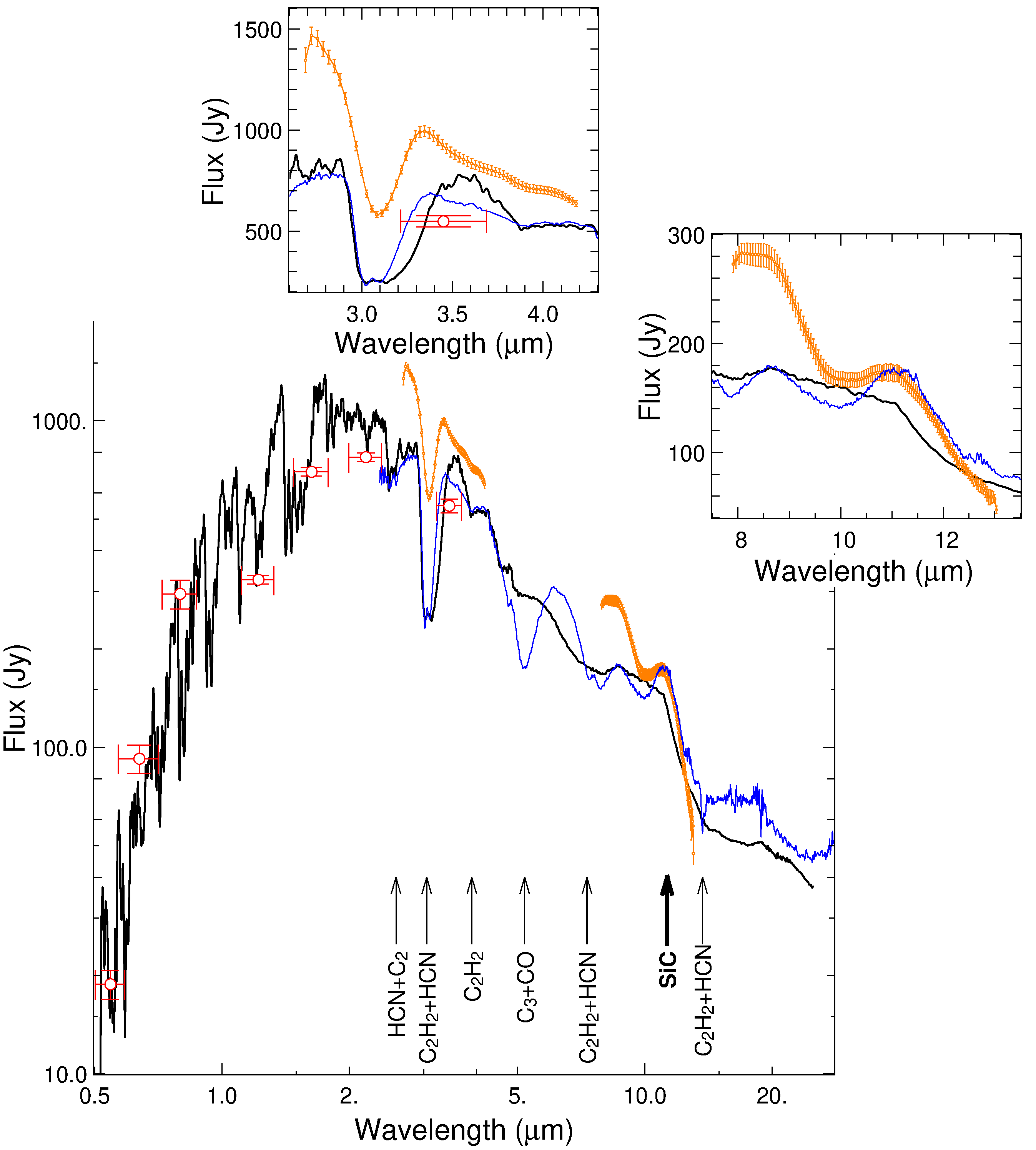}
\caption{SED of R~Scl (log-log scales). Orange spectra:~MATISSE data-sets in $L$- and $N$-bands with their associated error bars.  Blue spectra:~the \textit{ISO/SWS} spectrum. 
Red points:~photometric data from the literature. 
The solid black line is the best \texttt{COMARCS}+\texttt{DUSTY} model (see Section~\ref{sec:sed}). Spectral features are labeled with arrows. Insets zoom onto the MATISSE spectrum (linear scales). SiC is marked as boldface as it is solid state, unlike the other gaseous species.}
\label{fig:SED}
\end{center}
\end{figure}

\begin{figure*}[htbp]
\begin{center}
\includegraphics[width=1.0\textwidth]{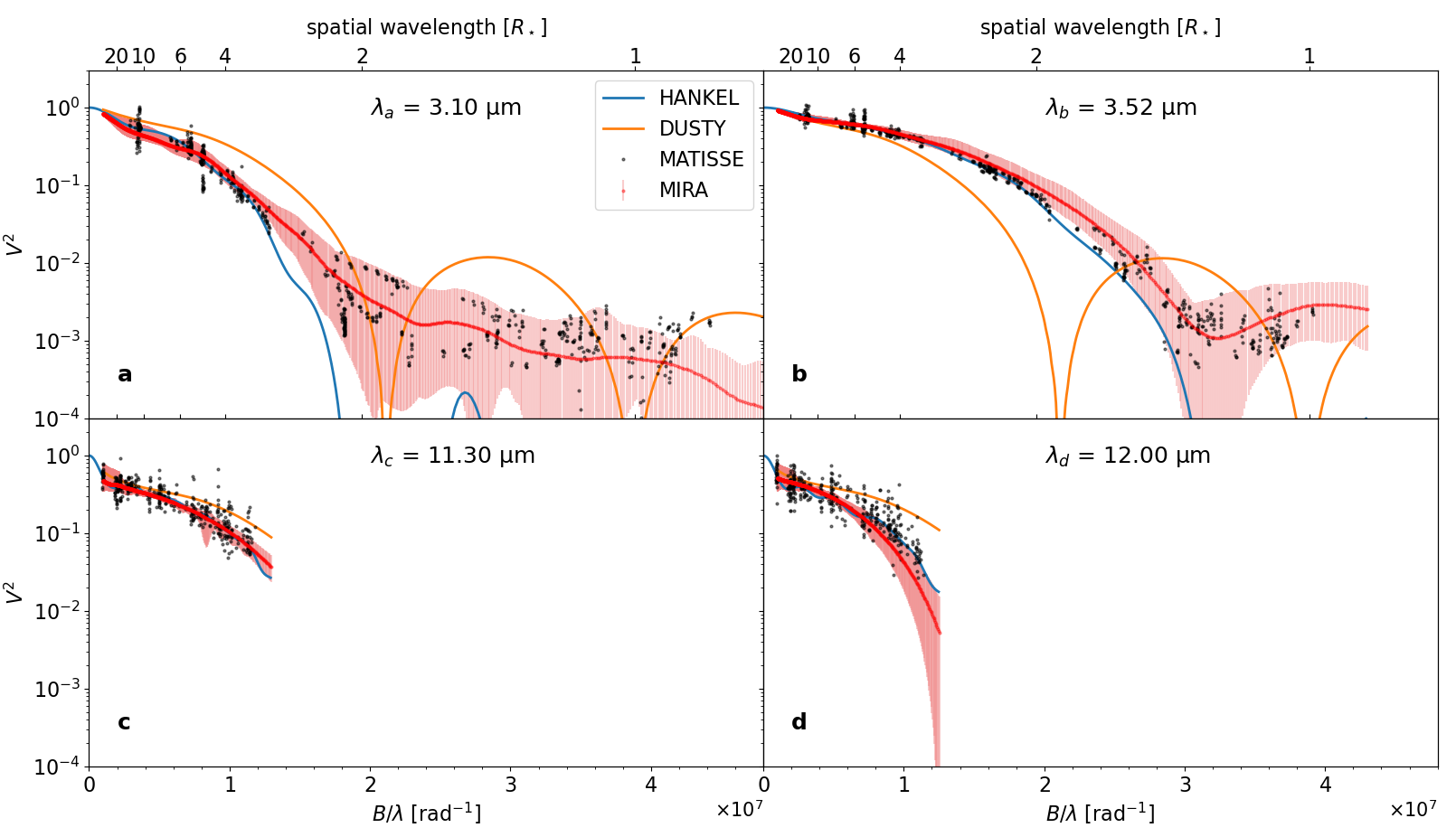}
\caption{Plot of the MATISSE squared calibrated visibilities (grey dots) at 3.10~$\mu m$  around the C$_2$H$_2$+HCN absorption feature (panel a), 3.52~$\mu m$ in the $L$-band pseudo-continuum (panel b), 11.30~$\mu m$ around the SiC feature (panel c), and 12.00~$\mu m$ in the $N$-band pseudo-continuum (panel d). We over-plotted the data with several models: Hankel profile (blue), \texttt{DUSTY} model (orange), and \texttt{MIRA} mean 1D radial profile (red) and its dispersion (solid red interval).}

\label{fig:V2_TOT}
\end{center}
\end{figure*}


The data are processed using the MATISSE data reduction software \texttt{drsmat} version 1.5.0 \citep{Millour2016}, which adopts a classical Fourier transforms scheme \citep{Perrin2003}, with photometric calibration in the case of the $L$-band \citep{Foresto1997}. Chopping correction (subtracting sky frames from target frames) and optical path difference (OPD) demodulation procedures are applied before summing up the spectral densities, bispectra, and interspectra, leading to estimates of the spectro-interferometric observables: spectral distributions in squared visibility, closure phase, and differential phase, in a similar way as was done with the AMBER instrument \citep{Tatulli2007}. This leads to four reduced OIFITS2 files per pointed target (one for each beam commutation), which contains as many observables as exposures on the target.

In the next step, an interferometric calibration procedure is applied to the data: the targets used as calibrators (CAL) are corrected from their expected observables given the predicted value of their angular diameters, and each science target (SCI) is corrected from this transfer function estimate, owing to a calibrated OIFITS2 file for each CAL-SCI pair\footnote{the calibrated data used in this work can be found at \url{http://oidb.jmmc.fr/collection.html?id=12803c68-6124-40ad-803c-68612470ad3d}}. It is important to notice that calibrators have been chosen with no IR excess (see \citet{cruzalebes2019} for a description of the IR excess-free criterion) to avoid bias on the infrared measurements and in the calibration process. The \texttt{drsmat} has been developed in C with the ESO-specific library CPL, and is interfaced with the ESO recipes environment \texttt{esorex}. A graphical user interface can be used in the form of python tools developed by the consortium\footnote{made freely available at \url{https://gitlab.oca.eu/MATISSE/tools/wikis/home}} to handle the data and recipes, and display the results.
Table~\ref{tab:calib} lists the stars used as calibrators for R~Scl during the MATISSE commissioning run.




\section{Matching hydrostatic photosphere and 1D dust envelope model with the SED} \label{sec:sed}



Following the approach of \cite{sacuto2011}, we performed an estimation of the spectral energy distribution (SED) of R~Scl at the pulsation phase of the MATISSE observations. Indeed, the SED is formed by selected photometric data points listed in Table~\ref{tab:photometric} and by the \textit{ISO/SWS} spectra convolved at the same spectral resolution as MATISSE ($R\sim30$) and chosen at a pulsation phase that is as close as possible to that of the MATISSE observations ($\varphi\sim0.95$, see Figure~\ref{fig:SED}).

\subsection{Stellar pulsation} \label{sec:pulsation-phase-section}

To estimate the pulsation phase of the MATISSE data, we made a periodogram analysis of the AAVSO $V$-band photometry. For this purpose, we downloaded data from between JD 2411972.3 and 2458893.5, and we selected only the best-quality data (based on the AAVSO data flag system). The periodogram not shown in this article is produced using the Lomb-Scargle method, an efficient algorithm for detecting and characterizing periodicity in unevenly-sampled time-data \citep{VanderPlas2018}. The primary peak was detected at $P = 372 \pm 5$ days which is consistent with the period 376\,days from \cite{wittkowski2017}. We also detected a secondary peak at 15000\,days ($\approx$ 41 years), which was previously unknown. We ensured this was not an alias of the shorter period by checking that the secondary peak remains after we removed the primary pulsation from the data. Furthermore, the 41-year pulsation period is consistent with long secondary periods detected in several AGB stars \citep{Wood1999, Soszy2021}. Henceforth, we refer  to the "pulsation" as the first 372-day period and "variability" as the long-term variability of 41~years. The zero-phase time is set at 2456245.5~days (=2012-11-14). The MATISSE observations were carried out close to maximum luminosity.


\subsection{Photometric data and dereddening}

The \textit{V}, \textit{R}, and \textit{I} magnitudes reported in Table~\ref{tab:photometric} are taken from the catalog of \citet{UBVRIJKL_catalog}. The errors on the flux are the results of errors in the magnitude, the color index, and the extinction. The \textit{J}, \textit{H}, \textit{K}, \textit{L} band magnitudes at the same pulsation phase as MATISSE data are taken from \citet{Whitelock2006}.

The photometric data are corrected from reddening using the visual extinction equation of \cite{Bagnulo1998}. To retrieve the extinction at each fiducial wavelength, we assumed a total-to-selective extinction ratio in the visible of $R = 3.1$ \citep{Fitzpatrick1999}. We notice that the dereddening effect induced by the interstellar medium on the observed photometric data points is small.

\subsection{Stellar contribution: COMARCS model fitting} \label{sec:comarcs-fitting}

In this section,  the stellar contribution will be described using hydrostatic models. The detailed fit of the spectral energy distribution which would require more advanced models such as dynamic ones \citep{Hoefner2005} is beyond the scope of the current paper. Our aim here is to derive the stellar parameters and get a hydrostatic model that will be used in the following section as central source for the radiative transfer equation solver \texttt{DUSTY} \footnote{made freely available at \url{http://faculty.washington.edu/ivezic/dusty_web/}} in order to model the total emission of R~Scl. \\
We fit the photometric data with the grid of precomputed low-resolution synthetic spectra \texttt{COMARCS} \citep{Aringer2016,Aringer2019}. The spectra are calculated starting from 1D hydrostatic model atmospheres that take into account atomic and molecular absorption directly computed using the opacity generation code  from the Copenhagen Opacities for Model Atmospheres (\texttt{COMA}) program \citep{Aringer2009}. \\
The fit is limited to the wavelength region where the dust contribution is negligible, namely, between  the \textit{V} and \textit{I}-bands (after dereddening). 
In such a wavelength domain, we could expect strong absorption from the circumstellar envelope (CSE) which is not considered in the models. Such dust extinction could, in principle, impact the effective temperature estimation of the \texttt{COMARCS} model. However, the bias described above is not significant for our purpose, as previously discussed in \cite{Brunner2018}. \\ 
Only a subset of 2600 \texttt{COMARCS} spectra matching metallicity, effective temperature, and stellar mass was selected from the original grid of 11248 models available in the official \texttt{COMARCS} repository\footnote{\url{http://stev.oapd.inaf.it/atm/lrspe.html}} \citep{Aringer2016}. After computing $\chi^2$ on the spectra subset, we selected one of the models within 3$\sigma$ of the minimum~$\chi^2$. \\
Using Equation 2 from \cite{Cruzalebes2013} we estimated the Rosseland radius associated to the model chosen in the previous step. Our result R$_\star = 5.3 \pm 0.6$ mas of the star is consistent with \cite{Cruzalebes2013} stellar limb-darkened radius of $R_{\star,C} = 5.03 \pm 0.03$ mas, based on hydrostatic \texttt{MARCS} model, and it is also consistent with the Rosseland radius $R_{\star,W} = 4.45 \pm 0.15$ mas of \cite{wittkowski2017}, obtained using dynamic atmosphere models. Even using different models, the latter two estimations are consistent within 1.5~$\sigma$ with our Rosseland radius, meaning that the differences are not statistically significant.  Finally, we also tested the \texttt{COMARCS} model against a distance of 393\,pc \citep{Gaia2018}, following the same steps than described previously, and we did not notice any significant change in the SED and the stellar contribution.

\subsection{Dust contribution: DUSTY model fitting} \label{sec:mass-loss-dust}

To describe the circumstellar envelope of the star, we use the 1D radiative transfer code \texttt{DUSTY}~\citep{Ivezic1997}. We set: (1) the central source as our best-fitting \texttt{COMARCS} model; (2) the dust grain-size distribution following the classical MRN (Mathis Rumpl Nordsieck) distribution \citep{Mathis1977}. For the latter, we assumed a minimum and maximum grain size $a_{min}$ = 0.005\,$\mu$m and $a_{max}$ = 0.25\,$\mu$m, respectively, which are largely adopted for size distribution among stellar envelopes and interstellar dust; (3) the inner shell temperature was set to 1200\,K, which aptly represents the sublimation-condensation temperature of dust in particular AmC \citep{Gail1999}. Such a temperature is affected, in practice, by unknowns related to dust grain optical and physical properties. As a consequence, we refer to the \texttt{DUSTY} radius where the dust grains condensate as the "\texttt{DUSTY} inner radius;" (4) then, we  arbitrarily set the outer radius of the envelope to 1000~\,$R_{in}$ consistently to \cite{sacuto2011}; (5) finally, we assumed the density distribution driven by the pressure on dust grains. Indeed, the Radiatively Driven Winds Analytic (RDWA) approximation \citep{Elitzur2001} was used for the dust density profile. The approach is appropriate for AGB stars (in most cases) and according to the \texttt{DUSTY} manual, it offers the advantage of a much shorter computing time. Then, fitting the model on both observed SED and the MATISSE visibilities, we finetuned the amount of silicone carbide (SiC) and amorphous carbon (amC), as well as the optical depth at 1\,$\mu$m. Thus, we estimated the inner shell radius and the mass-loss rate. We assume that the main error sources in the \texttt{DUSTY} fitting process come from the stellar temperature and luminosity. We used the lower and higher temperatures and luminosities of the \texttt{COMARCS} grid of models based on the error bars computed in  Sect.~\ref{sec:comarcs-fitting}. We repeated the process described in  Sect.~\ref{sec:mass-loss-dust} and deduced the confidence interval of the \texttt{DUSTY} parameters. The summary of the \texttt{DUSTY} parameters are given in Table~\ref{tab:res_TOT}.


\begin{table}[htbp]
\small
\begin{minipage}[t]{0.4\textwidth}
    \caption{Physico-chemical parameter values of the \texttt{COMARCS} (top part) and \texttt{DUSTY} (bottom part) models.}
    \centering
    \begin{tabular}{l | c | c} 
    Model parameter & Parameter value & Fixed \\ 
\hline \hline
    Distance & $360\pm50$\,pc \footnote{\citet{Maercker2018}} & Yes \\
    Surface gravity & $-0.50\pm0.10$\,cm.s$^{-2}$ & No \\
    Stellar surface temperature & $2700\pm100$\,K & No \\
    Stellar Luminosity & $8000\pm1000$\,$L_\odot$ & No \\
    Stellar mass & $2.0\pm0.5$\,$M_\odot$ & No \\
    C/O ratio & 2$^{+2.0}_{-0.6}$ & No \\
    Microturbulent velocity & 2.5\,km.s$^{-1}$ \footnote{\citet{Aringer2009}} & Yes \\
    Rosseland Radius & $1.91\pm0.20\,au$ & No \\ 
    &~ $5.3 \pm 0.6$ mas &  \\
    \hline
    Shell chemical composition & AmC \footnote{\citet{Hanner1988}} and SiC \footnote{\citet{Pegourie1988}} & No\\
    &~ $88\pm11$\%~of~AmC &   \\
    &~ $12\pm11$\%~of~SiC & \\
    Dust grain-size distribution & $n(a) \propto a^{-3.5}$ \footnote{\citet{Mathis1977}} & Yes \\
    Optical depth at $\lambda=1~\mu$m  & $0.19\pm0.05$ & No \\
    Inner boundary temperature & $1200\pm 100$\,K \footnote{\citet{sacuto2011}} & Yes \\
    Inner radius & $24.5\pm 9.0$\,mas & No \\
    &~ $4.6\pm 1.7 \,R_\star$ &  \\
    Outer radius & 1000\,$R_{in}$ $^e$ & Yes \\
    Mass-loss rate & $ 1.2\pm 0.4 \times 10^{-6} M_{\odot}~\rm{yr}^{-1}$ & No \\
\hline \hline
\end{tabular}
    \label{tab:res_TOT}
\end{minipage}
\end{table}


\subsection{Results of the composite model fitting}

Table~\ref{tab:res_TOT} summarizes the parameters of the composite \texttt{COMARCS}+\texttt{DUSTY} best-fit model. The fractional abundance of the relevant grains in the dust shell resulting from this fit is a mixture of amorphous carbon (AmC, 88 $\pm$ 11\%) and silicon carbide (SiC, 12 $\pm$ 11\%), with the optical depth at $1 \mu$m $\tau_{\lambda = 1 \mu m} = 0.19 \pm 0.05$. Our results are consistent with the findings of \cite{sacuto2011}, who used a grid of \texttt{DUSTY} models and least-squares fitting minimization to find a mixture of 90 $\pm$ 10\% AmC and 10 $\pm$ 10\% of SiC and an optical depth at $1~\mu$m of $\tau_{\lambda = 1 \mu m} = 0.18 \pm 0.05$. Then, we also estimated a mass-loss rate of $\dot M = (1.2\pm 0.4) \times 10^{-6} M_{\odot}~\rm{yr}^{-1}$, using a dust grain bulk density $\rho_s=1.85\,\rm{g.cm}^{-3}$ \citep{Rouleau1991} and a gas-to-dust-ratio of $r_{gd} = 590$ \citep{Schoier2005}. Our estimation of the mass-loss rate is also consistent with that of \cite{deBeck2010}, namely,  $1.6 \times 10^{-6} M_{\odot}~\rm{yr}^{-1}$.

In Fig.~\ref{fig:SED}, we display the selected photometry data sets and the composite model of the SED. The MATISSE and the \textit{ISO/SWS} spectra clearly show signatures  possibly associated to acetylene (C$_2$H$_2$) and hydrogen cyanide (HCN) molecules at 3.1\,$\mu$m as well as the solid state SiC at 11.3\,$\mu$m \citep{Yang2004}. We suspect the vertical-axis offset of the \texttt{MATISSE} spectrum with respect to the ISO observations and \texttt{COMARCS+DUSTY} models is due to calibration issues in terms of the \texttt{MATISSE} absolute flux or a field-of-view effect \citep{paladini2017}. We note that \texttt{COMARCS + DUSTY} models cannot reproduce properly the CO + C$_3$ absorption feature at 5.2 \,$\mu$m. Part of this discrepancy can be linked to a well known problem about the opacity data and equilibrium constants involved in the C$_3$ band chemical description \citep{Aringer2019}.


Since \texttt{DUSTY} does not handle molecules, and knowing the coexistence of molecules and dust, we expected to see strong discrepancies between the \texttt{DUSTY} model and the MATISSE data in the $L$-band, where strong molecular features are located. Although dust emission usually dominates the $N$-band, molecular contributions are also expected there, namely: CS and possibly SiO at 8\,$\mu$m; HCN+C$_2$H$_2$ molecular bands between 11 and 16\,$\mu$m \citep{Chubb2020} which can affect the shape of the SiC band at 11.3\,$\mu$m.

The resulting fit to the SED (black curve in Fig.~\ref{fig:SED}) provides results that are in reasonable agreement with the above expectations: a satisfactory fit to the $N$-band visibilities from MATISSE is found (orange curves in Fig.~\ref{fig:V2_TOT} c and d), but the $L$-band visibilities are not well matched with this star+dust model alone (Fig.~\ref{fig:V2_TOT} a and b). We therefore need to refine this model, introducing additional ingredients for the $L$-band.

\section{R~Scl as seen by MATISSE} \label{sec:model}


To locate the dusty and molecular regions, we used two independent methods: the first one based on the \texttt{MIRA} image reconstruction algorithm and the second one a new approach that is presented in this paper and based on an intensity radial profile reconstruction method using the Hankel transform. The Hankel method has the advantage of requiring less parameters than image reconstruction, leading to fewer solution non-uniqueness problems, a better convergence, and a better dynamic range, given that the source is centro-symmetric at approximately all wavelengths.

\subsection{MIRA image reconstruction} \label{sec:image}

In Appendix.~\ref{fig:uvcov}, we show the (u,v)-plane coverage obtained during the observing run. The coverage is very dense, allowing us to reconstruct high-fidelity monochromatic images of R~Scl. 
For that purpose, we fed the calibrated visibilities and closure phases into the \texttt{MIRA} software of \citet{Thiebaut2008}. Additional scripts written by \citet{Millour2012} allowed us to slice the wavelengths, and improve the image quality by low-frequencies filling.

\begin{figure*}[htbp]
\begin{center}
\includegraphics[width=0.98\textwidth]{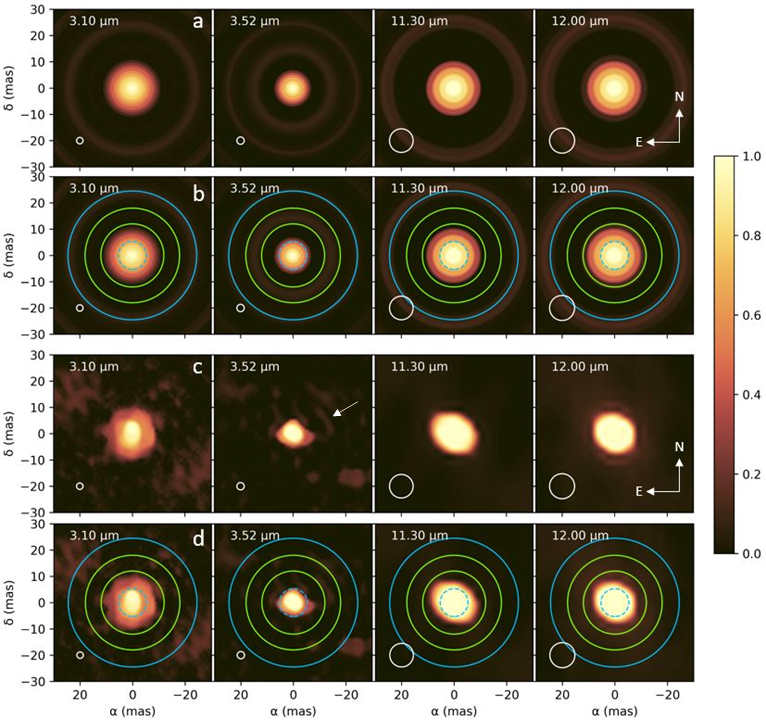}
\caption{Reconstructed Hankel distribution from radial profiles (row a), the same with the identified features highlighted (row b), images reconstructed with \texttt{MIRA} (row c), and the same with features highlighted (row d). Each panel shows the reconstructed image at the wavelength shown at its left top corner, covering the $L$-band (3--4\,$\mu$m) and the $N$-band (8--13\,$\mu$m). In these panels, the blue dashed circle in the center shows the calculated Rosseland radius of the stellar photosphere, the green circles represent the inner and outer boundary of the distinct molecular shell (seen here only at 3.52\,$\mu$m and only clearly identifiable in the Hankel reconstruction), the blue circle represents the inner boundary of the dust envelope predicted by DUSTY, while the white circle at each bottom left corner shows the theoretical angular resolution of the interferometer. The white arrow shows the north-west arc structure referenced in Sect.~\ref{sec:distctlayer}.}
\label{fig:images}
\end{center}
\end{figure*}

The images were reconstructed with 128 pixel size, using total variation regularization, a hyperparameter value of 10,000 \citep[as suggested in][]{2011Renard}, a random start image for the first wavelength, and the median image of all previous wavelengths for the subsequent ones.
In rows c and d of Fig.~\ref{fig:images}, we show a subset of the reconstructed images. 

Since the reconstructed interferometric image has a finer resolution than the diffraction-limited resolution, we also show in Fig.~\ref{fig:images_convol} the reconstructed images convolved by an interferometer clean beam with a Gaussian profile. Furthermore, our \texttt{MIRA} image reconstruction effectively reproduces the closure phase of MATISSE data at all wavelengths -- even at 3.11\,$\mu$m, as shown in Appendix~\ref{fig:closurephase}, where strong asymmetries are observed.

\subsection{Hankel Profile reconstruction}\label{sec:hankel}

The Hankel profile method uses a large number of concentric uniform narrow rings of same width as that given by the angular resolution of the instrument. A detailed description of the \texttt{RHAPSODY} tool is given in Appendix~\ref{sec:Hankelprofile}. Although it is built on simple models, \texttt{RHAPSODY} is not a simple model-fitting tool; rather, it is a radial profile reconstruction tool based on a Bayesian approach.

\begin{figure}[htbp]
\includegraphics[width=0.48\textwidth]{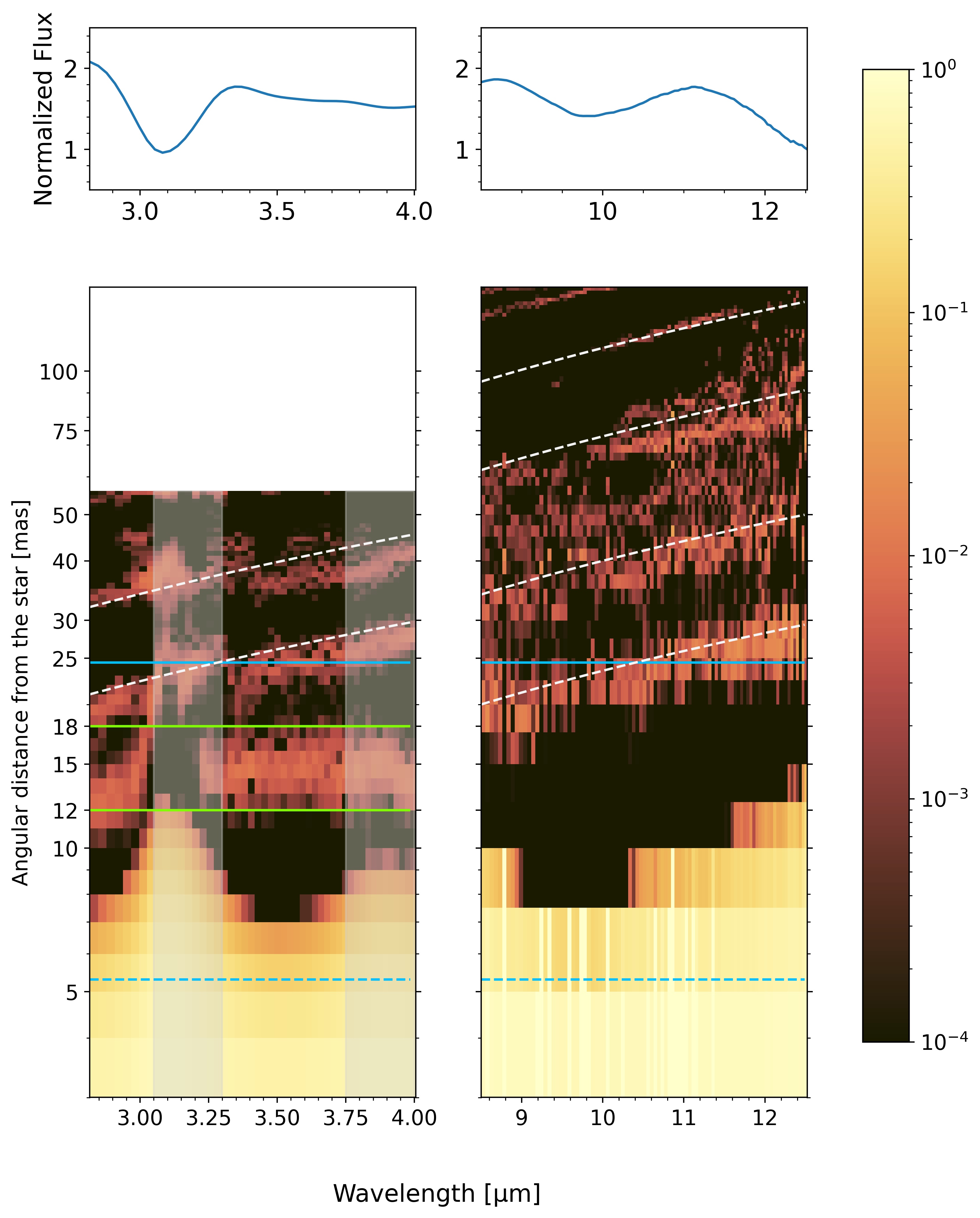}
\caption{Spectra of the intensity profiles in $L$- and $N$-bands. Upper panels correspond to the observed MATISSE spectrum normalized by the median and then divided by a black body spectra at T=2700\,K to underline the observed emission and absorption features. Lower panels are the spectro-radial maps in $L$- and $N$-band obtained by plotting the best intensity Hankel profile normalized at one for each observed wavelength. The faint red structures highlighted with the inclined dashed-white lines are reconstruction artifacts (see Sect.~\ref{sec:hankel} for details). The dashed and solid blue horizontal lines show the position of the Rosseland radius and the \texttt{DUSTY} inner radius respectively. The solid green horizontal lines delimit the extension of a hot distinct molecular layer (see Sect.~\ref{sec:molecular}). The grey vertical bands cover the spectral ranges where the centro-symmetric Hankel profile is not able to properly reproduce the asymmetric shape of R~Scl revealed by non-zero closure phases.}
\label{fig:spectro_radial_intens}
\end{figure}

In this work, we reconstructed monochromatic radial profiles from visibilities using this technique. Each radial profiles is generated varying several parameters, including the number of rings, $N_{\rm step}$, the number of iterations in the minimization routine, the initial radial profile, and the value of the hyperparameter $\mu$ that we varied between $\mu=1$ and $\mu=10^{10}$, in increments of the power of ten. The L-curve ($\chi^2$ as a function of $\mu$) provides us a guideline to select the optimal value using both regularization methods: 1) the quadratic smoothness $f_{\rm smooth}$, which tends to minimize the intensity variations between consecutive radii; and 2) the total variation $f_{\rm var}$ which tends to reconstruct uniform intensities with sharp edges. Then we obtain the best radial profile reconstructions (shown in Figs.~\ref{fig:images} and \ref{fig:spectro_radial_intens}) with $\mu = 10^4$ and a median value of the minimum reduced $\chi^2$ of 3.3, using the \texttt{total variation} regularization method.

Images of the reconstructed Hankel profiles at selected wavelengths (i.e., pseudo-continuum (3.52\,$\mu$m) and in C$_2$H$_2$ and/or HCN absorption bands (3.10\,$\mu$m, >\,12\,$\mu$m), as well as the SiC signature (11.30\,$\mu$m)) are shown in Fig.~\ref{fig:images}. The estimated Rosseland diameter is marked as a dashed circle and the typical angular resolution of the interferometer for the fiducial wavelength, computed as $0.5\,\lambda/B$ \citep{Monnier2003}, is marked as a white circle in the bottom-left corner.


In Fig.~\ref{fig:spectro_radial_intens}, we show the obtained spectro-radial maps of the Hankel profile in the $L$- and $N$-bands. Profile reconstruction artifacts, coming from the (u,v)-plane sampling voids, are clearly identifiable as their distance from the star depends on the wavelength across the two $L$- and $N$-bands (traced as dashed white lines in Fig.~\ref{fig:spectro_radial_intens}). Spatially well-constrained structures show no dependence of the distance on wavelength ($ d = \text{Const}[\lambda]$). We refer to Appendix~\ref{sec:rhapsody_detection} for more details on the fidelity of the profiles. The two grey vertical bands from 3.05 to 3.30\,$\mu$m and from 3.75 to 4.00\,$\mu$m cover the spectral ranges where the centro-symmetric Hankel profile is not able to properly reproduce the asymmetric shape of R~Scl, revealed by non-zero closure phases and by the reconstructed images (see the following section). The Rosseland radius is represented as a dashed blue line, the \texttt{DUSTY} inner radius in solid blue line, and a new, so far unexpected emission discussed in Section~\ref{sec:molecular}, which is between 12 and 18\,mas~($\sim$ 2.3--3.4\,$R_\star$) within the two solid green lines.

As seen in Fig.~\ref{fig:spectro_radial_intens}, the extent of the inmost circumstellar shell is not identical at all wavelengths but this variation is not coming from any instrumental or data reduction issue. The shell is significantly larger between 2.9\,$\mu$m and 3.3\,$\mu$m, and also 3.6\,$\mu$m and 4.0\,$\mu$m, and except for those specific bandwidths, it steadily increases towards longer wavelengths all the way up to 12\,$\mu$m. Furthermore, this is in good agreement with \cite{Paladini2009}: from dynamical model, the pseudo continuum radius is larger than the Rosseland radius because of the atmospheric extension and the pseudo-continuous molecular opacity. Notably, even at 3.525\,$\mu$m, where we are expecting to find only the pseudo-continuum \citep{Paladini2009}, the radial profile extent is slightly larger than the value of the Rosseland radius. Therefore, our observations are qualitatively supportive of the synthetic observations derived on the basis of dynamic models described in \cite{Paladini2009}.

The two reconstruction processes (profile and image) are different and, therefore, they produce different results that must be interpreted adequately: the Hankel profiles are computed using the visibilities across all position angles altogether, reconstructing only the centro-symmetric information of the object, whereas the \texttt{MIRA} images are computed using the visibilities and the closure phases as a function of position angles and they bear the asymmetry information. As a consequence, it is not surprising to find asymmetric structures in the MIRA images while not seeing these asymmetries in the Hankel profiles. Therefore, we should not go too far in the comparison of images coming from both methods: what can be compared are the radial positions of features, but not their relative intensities (except for azimuthally integrated profiles of \texttt{MIRA} images compared to the Hankel profiles shown in Fig.~\ref{fig:images_spectra} and described in the following subsection). Furthermore, since it is difficult to estimate the true angular resolution and the shape of the PSF, the pixel size used to reconstruct the images is not meant to be the exact angular resolution of the image for both methods.

\subsection{Description of the R Scl environment in the mid-IR} \label{sec:description}

\begin{figure}[htbp]
\begin{center}
\includegraphics[width=0.48\textwidth]{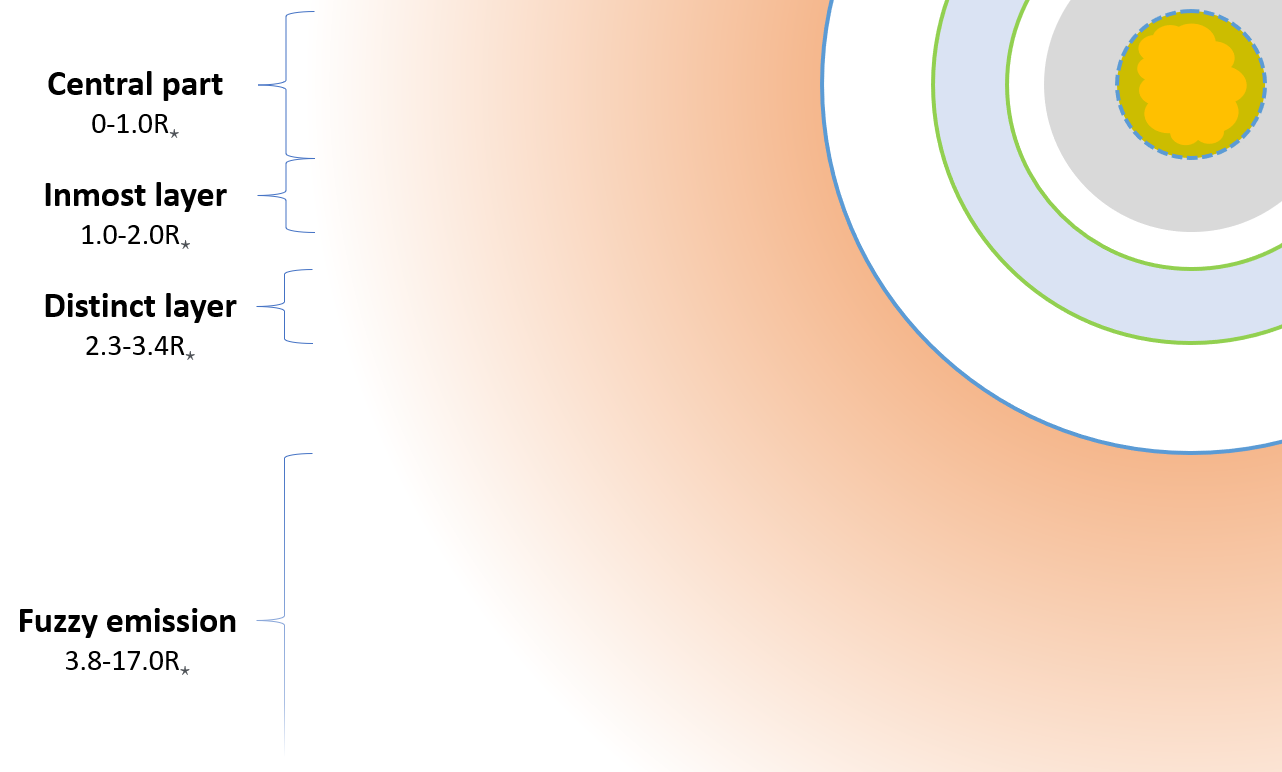}
\caption{Shell structure detected in the close envelope of R Scl (stellar radii shown ranges).}
\label{fig:global}
\end{center}
\end{figure}

In this section, we describe the observations, while the interpretations can be found in Sect.~\ref{sec:discussion}.
In Fig.~\ref{fig:global}, we provide a sketch of the region of the circumstellar envelope close to the star R~Scl based on the Hankel profiles and the MIRA images, which we describe in more detail in the following sections. Such a circumstellar envelope is mainly composed of: i) the central component from 0 to 5 mas from the center of the star; ii) an inmost layer from 5 to 10 mas~(i.e.,~$\sim$ 1--2\,$R_\star$); iii) a distinct layer from 12 to 18 mas~(i.e.,~$\sim$ 2.3--3.4\,$R_\star$); iv) a fuzzy emission in the stellar outskirts at 20 to 90 mas~(i.e.,~$\sim$ 3.8--17.0\,$R_\star$).

\subsubsection{Central component (0 -- 5\,mas)}

In the first region between 0 and 5 mas from the center of the star (Figs.~\ref{fig:images} and \ref{fig:spectro_radial_intens}), we see a nearly round source dominating the flux in the pseudo-continuum (3.52\,$\mu$m and 10\,$\mu$m). Figure~\ref{fig:images_spectra} shows spectra extracted from the Hankel profiles and the MIRA images in different regions around the star. These separated spectra allow us to better understand the nature and the composition of the different layer around the star. In Fig.~\ref{fig:images_spectra}, we see that the spectrum of the central source is very similar to the total one: an absorption feature at 3.05\,$\mu$m and two bumps in the $N$-band at 8.7\,$\mu$m and around 11.1\,$\mu$m. This central part is clearly asymmetric in the 3.1\,$\mu$m C$_2$H$_2$/HCN absorption band in the \texttt{MIRA} images.

\subsubsection{Inmost layer (5 -- 10\,mas)}

A layer beyond the estimated Rosseland radius of R~Scl is seen as a faint environment between 5 and 7\,mas ($\sim$ 1.0--1.3\,$R_\star$), assuming that the border of the photosphere is at 5\,mas from the center of the star, as also seen in the images in Fig.~\ref{fig:images}. The $N$-band images also exhibit a larger central object compared to the $L$-band central star, meaning that we see the contributions from the star and the inmost layer, but we cannot disentangle them due to the lower angular resolution at these longer wavelengths.

We also observe a similar structure extending up to 10\,mas (1.8 $R_\star$) and attached to the central component, visible in the MIRA images at 3.1\,$\mu$m and 3.9\,$\mu$m as an irregular extended layer attached to the central component (Fig.~\ref{fig:images}) and as a ring around the central object in the Hankel profiles. In the spectra of Fig.~\ref{fig:images_spectra}, this is seen as two emissions bumps at 3.05\,$\mu$m and 3.9\,$\mu$m.

\subsubsection{Distinct layer (12 -- 18\,mas)} \label{sec:distctlayer}

A layer that is apparently distinct from the previous one (i.e., continuum at 3.52\,$\mu$m) is detected between 12 and 18\,mas, below the \texttt{DUSTY} inner radius of 24.5\,mas. Such layer is seen in the $L$-band in the \texttt{MIRA} image reconstruction (Figs.~\ref{fig:images} and \ref{fig:images_contour}) as a fuzzy arc to the north-west of the star and a few blobs at the same distance from the photosphere. The \texttt{RHAPSODY} intensity profile reconstruction shows also such layer as a continuous ring. The nature of such layer is further discussed in Sect.\ref{sec:molecular}.

\subsubsection{Fuzzy emission in the outer skirts (beyond 20\,mas)}

A fuzzy emission above 20\, mas (marked as a solid blue line in Figs.~\ref{fig:images} and \ref{fig:spectro_radial_intens}) is detected with MIRA and RHAPSODY. This emission is present both in $L$- and $N$-band. In the RHAPSODY images, the dynamic range of this emission is so low that only flux artifacts are reconstructed (see Appendix~\ref{sec:rhapsody_detection}). In the MIRA images, the (u,v)-coverage does not allow us to meaningfully reconstruct the shape of the brightness distribution. As a consequence, the fuzzy emission appears as blobs or artifacts in the north-eastern and south-western directions.

\section{Discussion}
\label{sec:discussion}

In this paper, we describe 2D images that are projections of the actual R~Scl 3D environment. Consequently, the distances mentioned in the following sections refer by default to distances projected onto the plane of the sky.

\begin{figure}[htbp]
\begin{center}
\includegraphics[width=0.48\textwidth]{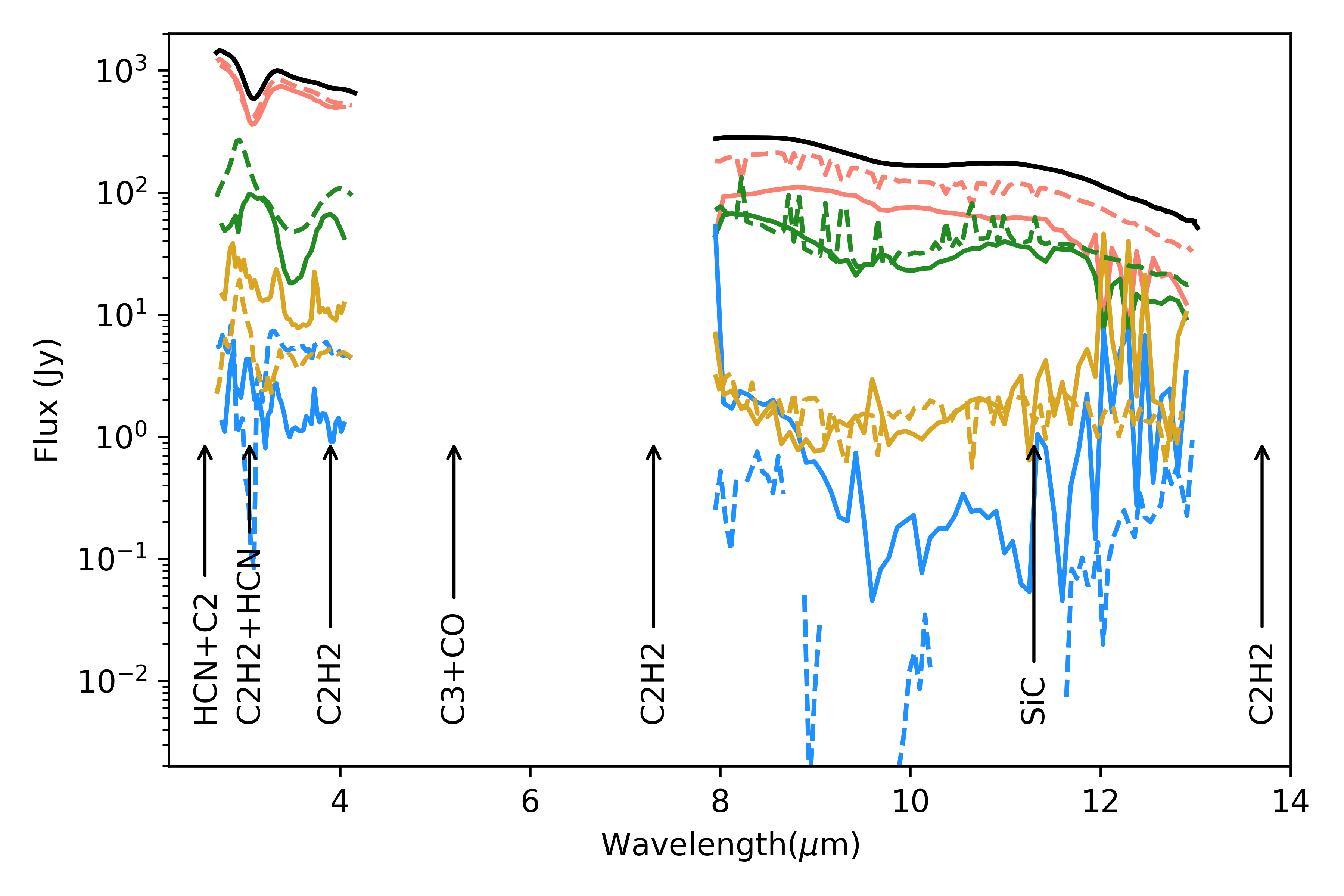}
\caption{MATISSE spectrum (black line) together with spectra extracted from circumstellar regions located at 0--5\,mas (pink), 5--10\,mas (green), 12--18\,mas (blue), and 20--90\,mas (golden) for both the Hankel (solid lines) and \texttt{MIRA} reconstructed images (dashed lines).}
\label{fig:images_spectra}
\end{center}
\end{figure}


\subsection{Central part: Star veiled by molecules}

The data show a structure between 0--5\,mas which can be interpreted as the star's photosphere. At the wavelengths where molecular bands are expected, the photosphere is no longer visible due to the presence of molecular shells in front of it.
In the $L$-band at 3.05\,$\mu$m, we can see a spectral feature which, according to, for instance, \cite{Aringer2019}, could be attributed to C$_2$H$_2$ + HCN. The feature detected at 3.9\,$\mu$m is associated to C$_2$H$_2$.

The bump near 11.3\,$\mu$m in the $N$-band (see Figs.~\ref{fig:SED} and \ref{fig:images_spectra}) resembles the SiC signature. Two hypothesis are able to explain such feature: i) The signature is not from dust but rather from a pseudo-continuum between adjacent molecular features; ii) The signature is produced by an optically thick layer of SiC dust located along the line of sight. In the case where the layer temperature is lower than the one of the background (i.e., the stellar surface), an optically thin layer scenario would produce an absorption \citep{Rutten2000}; hence, this would not be compatible with the emission observed. The limited angular resolution and the dynamic range of the MATISSE $N$-band data do not allow us to correctly probe the SiC emission region, which provides key information in making the choice of one scenario over the other.


\subsection{Inmost hot molecular layer (5--10 mas)}

The inmost layer has clear spectral features (seen in the green curve of Fig.~\ref{fig:images_spectra}), appearing as two emission bumps at 3.05\,$\mu$m and 3.9\,$\mu$m, namely, the signatures of C$_2$H$_2$ and HCN. The hypothesis of the existence of an absorption band between these bumps is unlikely because spectral features at 3.05\,$\mu$m and 3.9\,$\mu$m should therefore also be in absorption. In addition, the data angular resolution in the $L$-band is high enough to avoid any contamination from the stellar emission. Furthermore, the existence of these emission bands is consistent with the lack of a strong continuum emission coming from the back hemisphere of the envelope, in the ring between 5 and 10\,mas. On the other hand, C$_2$H$_2$ and HCN are already formed in the stellar atmosphere at a temperature between 2500\,K and 3000\,K \citep{Eriksson1984}; thus, we suggest that this layer is attached to the star and could be part of the extended stellar atmosphere.

The inmost hot molecular layer is also visible at all other wavelengths in the $L$-band as a faint environment between 5 and 7\,mas ($\sim$ 1.0--1.3\,$R_\star$), as seen in the images in Fig.~\ref{fig:images}. It may also be seen in the $N$-band, as the images show a larger central object with respect to the photosphere of the star.

\subsection{Possibility of a distinct hot molecular layer (12--18 mas)}  \label{sec:molecular}

In $L$-band panels of Fig.~\ref{fig:images}, the \texttt{MIRA} images indicate the presence of irregular emissions between 12 and 18 mas while \texttt{RHAPSODY} images (as also seen in Fig.~\ref{fig:spectro_radial_intens}) shows clearly the presence of a layer in emission.
In contrast, in the $N$-band, neither \texttt{MIRA} nor \texttt{RHAPSODY} show any trace of such layer, but there is still a noticeable enlargement with regard to the angular diameter of the photosphere. The non-detection of such layer might be induced by the limited dynamic range and $N$-band angular resolution of the data (see Appendix~\ref{sec:rhapsody_detection}).
The composition of this layer can then be of two different natures: either formed by dust or filled with molecules.

\subsubsection{Dust hypothesis}\label{sec:dusthyp}

 The dust hypothesis, based on the assumption the 12--18\,mas layer (2.3--3.4 $R_\star$) is formed by dust, offers three possibilities: i) Dust grains in the layer are only made of amorphous carbon. According to our estimation using Equation 1 in \cite{Ivezic1996}, such layer should radiate at a mean temperature of 1500\,K. This temperature is in accordance with the condensation temperature of amC in dynamic models \citep{Nowotny2010}. \cite{wittkowski2017} also confirmed the presence of amC in the line of sight of their R~Scl PIONIER images; ii) Dust grains in the layer are made of SiC. If we assume, as the temperature for the SiC dust condensation, T$_{eff}$ = 1350\,K from \cite{Yasuda2012}, then such a scenario would be discarded because the estimated temperature of the layer in this work is higher than the latter. On the other hand, we should keep in mind that the condensation temperature of SiC is not well known. \cite{Yasuda2012} used select data on the condensation of Si$_x$C$_y$ onto SiC dust grains that are affected by large errors. \cite{Gobrecht2017} analyzed part of the chemical reactions used by \cite{Yasuda2012} and found significant differences. \cite{Agundez2020} gave a condensation temperature for SiC of $\approx$\,1300--1400\,K, while \cite{Menshchikov2001} provided a temperature of 2000\,K. Given these uncertainties, the scenario cannot be fully discarded; iii) Dust grains in the layer are made of titanium carbide (TiC). TiC also condenses at those temperatures \citep{Agundez2020}, however such a chemical species is not included in this paper since there is no clear spectral signature of TiC in the MATISSE spectral range \citep{Helden2000}.

\begin{figure*}[htbp]
\begin{center}
\includegraphics[width=0.9\textwidth]{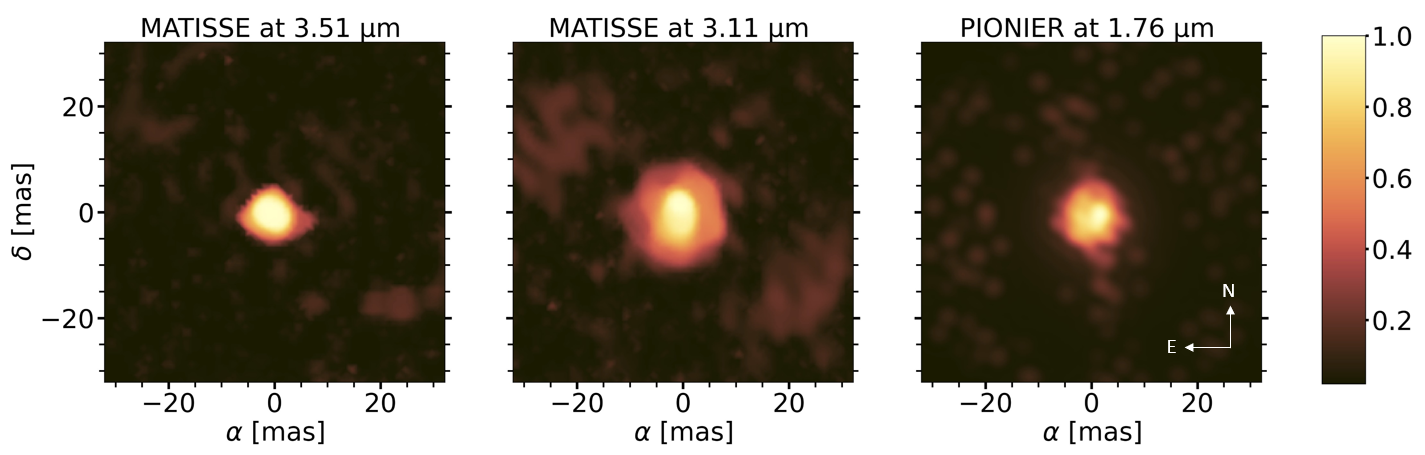}
\caption{Comparison between R~Scl in 2014 and 2018 observed respectively with PIONIER and MATISSE. Left: MIRA image reconstruction at 3.51\,$\mu$m using the 2018 MATISSE observations at $\phi$ = 0.96. Center: MIRA image reconstruction at 3.11\,$\mu$m using the 2018 MATISSE observations at $\phi$ = 0.96. Right: IRBis image reconstruction made by \cite{wittkowski2017} using PIONIER 2014 data at 1.76\,$\mu$m and $\phi$ = 0.78.}
\label{fig:wittkowski}
\end{center}
\end{figure*}

\begin{table*}[htbp]
    \caption{Summary of key structures in the inner region of R Scl from our MATISSE observations.}
    \centering
    \begin{tabular}{l | c | c | c | c} 
    Structure & Composition & Distance from the stellar center & Spectral signature & Central Wavelength \\ 
\hline \hline
    Star    & -- & $0- (5.3\pm0.6)$,mas &\ Continuum & All \\
    Rosseland Radius & --  & $5.3\pm0.6$\,mas~($\sim$~0--1~$R_\star$) & -- & --   \\
    Central Component & C$_2$H$_2$ + HCN & $0-5$\,mas~($\sim$~0--1~$R_\star$) & Absorption band & 3.05\,$\mu$m   \\
    Inmost Layer & C$_2$H$_2$ + HCN & $5-10$\,mas~($\sim$~1--2~$R_\star$) &  Emission band &  3.05 and 3.9\,$\mu$m \\
    Distinct Layer & C$_2$H$_2$ + HCN + amC & $12-18$\,mas~($\sim$~2.3--3.4~$R_\star$)  & Not clear & -- \\
    Fuzzy emission & amC (and SiC?) & $20-90$\,mas~($\sim$~3.8--17.0~$R_\star$) & Continuum & All \\
    \hline
\hline \hline
\end{tabular}
    \label{tab:summary}
\end{table*}

. 

\subsubsection{Molecular hypothesis} \label{sec:molhyp}

In the case of a molecular layer, it might be formed by C$_2$H$_2$ + HCN molecules. Indeed, in the ISO/SWS spectra of Fig.~\ref{fig:SED}, we observe a minimum around 10 \,$\mu$m in $N$-band, which could be the pseudo-continuum between C$_2$H$_2$ + HCN features -- at 7.5\,$\mu$m [6.5\,$\mu$m--9.0\,$\mu$m] and 13.0\,$\mu$m [11.0\,$\mu$m--16.0\,$\mu$m] \citealt{Chubb2020}. Such a possible pseudo-continuum would be bordered at long wavelengths by the SiC band. Furthermore, we should also notice that weaker molecular bands as SiH$_4$, NH$_3$, or C$_2$H$_4$ might also contribute to the shape of the SED at those wavelengths. On the other hand, similar layers were detected around AGB stars in previous works: \object{HD 187796} ($\chi$~Cyg) was observed by \cite{perrin2004} with a molecular layer detected between 1.8-1.9\,$R_\star$, \cite{zhao2012} revealed a molecular layer made of C$_2$H$_2$ and HCN at about 2\,$R_\star$ around the carbon rich star \object{V Hya}. Similar sizes can be found in oxygen-rich AGB stars \citep[BK~Vir: 1.2-4.5\,$R_\star$, SW~Vir: 1.2-3.0\,$R_\star$ in ][]{Hadjara2019}. \\
The molecule detection at this distance from the star echoes the results of \cite{wittkowski2017} which modeled AMBER and PIONIER data of R~Scl using \texttt{COMA} showing evidence for molecular contributions by CO, CN, C$_2$, C$_2$H$_2$, HCN, and C$_3$ close to the star. It is worthwhile to note in this respect that PAHs and very small grains seem to show broad emission plateaus in the $N$-band which would produce a similar spectral signature \citep{Peters2017}. Such particles belong to the formation process of amC and would thus be present in the dust condensation region.\\
\\

Following the reasoning presented in Sect.~\ref{sec:dusthyp} and Sect.~\ref{sec:molhyp}, it is very likely that the 12--18\,mas layers are composed by C$_2$H$_2$, HCN, and (possibly) amC dust.

\subsection{Comparison with PIONIER data}

In Figure~\ref{fig:wittkowski}, we compare our VLTI/MATISSE images with those from VLTI/PIONIER from \cite{wittkowski2017} \footnote{\texttt{IRBis} and \texttt{MIRA} packages are based on the same principle, which justifies the comparison.}.

The PIONIER 1.76\,$\mu$m observations give us information mainly on the stellar surface and its extended atmosphere, being outside of the 1.53\,$\mu$m C$_2$H$_2$+HCN molecular absorption band. We need to keep in mind that C$_2$ bands are also present in the observing bandwidth of PIONIER and could affect the images of the continuum emission.
The 1.76\,$\mu$m VLTI/PIONER data were observed in 2014, corresponding to a pulsation phase of $\phi = 0.78$, and the VLTI/MATISSE data were observed in 2018 at a pulsation phase of $\phi = 0.96$. The two epochs at four years apart are expected to show an evolution: the long-term variability of the star had time to completely reshape the region of the envelope close to the inmost layers of the vicinity of the star.
The comparison shows similarities: the size of the central source (0--5 mas) with PIONIER is akin to our central source at 3.51\,$\mu$m (in the pseudo-continuum).


It also exhibits some differences: asymmetries are not located at the same place between the two epochs. Two hypotheses for this variation: a significant evolution of the stellar and circumstellar environment shape or wavelength-dependent opacities related to the presence of molecules and possibly dust in front of the star. 
Then, this comparison has underlined the fact that it is possible to observe stellar surface variability at a timescale of few years for this kind of object. This means that in the case of a high observation frequency with MATISSE or another imaging interferometer, we should be able to estimate the characteristic timescale and sizes of patterns caused by convection onto the stellar surface.

\section{Conclusion}\label{sec:conclusion}

In this paper, for the first time, we present our processing of simultaneous $L$-band and $N$-band observational data of the AGB star R~Scl obtained with the VLTI/MATISSE instrument. Thanks to the good (u,v)-plane coverage of the MATISSE observations, we derived a consistent \texttt{MIRA} image and Hankel intensity profile reconstruction. The use of both independent and complementary methods allows us to locate the dust and molecules around R~Scl. In the images (and intensity profiles), we find the signature of a sporadic mass-loss event maybe responsible for the formation of a distinct layer. Based on the \texttt{MIRA} images, we also detected some asymmetric structures confirming the presence of clumps in the envelope of R~Scl. We did not find any clear signature of SiC forming regions around the star.
We present the main steps of our work as follows:

   \begin{enumerate}
      \item Data obtained with the VLTI/MATISSE instrument during the seven-days of the long commissioning session from 2018-12-03~to~2018-12-15 was used in this work.
      \item Using the \texttt{COMARCS} stellar atmospheric model and the dust radiative transfer code \texttt{DUSTY,} we confirm the previously published characteristics of a thin dust shell, mainly composed of amorphous carbon with a small amount of silicon carbide (10\%), leading to the location of the inner boundary estimated by \texttt{DUSTY} at about 4.6\,$R_\star$ from the central star (i.e., 24.4 mas) and a mass-loss rate of $ 1.2\pm 0.4 \times 10^{-6} M_{\odot}~\rm{yr}^{-1}$.
      \item In order to locate the dust and molecule regions, we compared two different methods: one based on the \texttt{MIRA} image reconstruction and an other one based on a brand-new approach which consists of reconstructing the Hankel intensity profile of the circumstellar environment using the code \texttt{RHAPSODY}\footnote{\href{https://github.com/jdrevon/RHAPSODY}{Link to RHAPSODY: https://github.com/jdrevon/RHAPSODY}}. This method allowed us to simultaneously reproduce the spectral energy distribution and the MATISSE visibilities.
      \item These complementary and independent approaches allow us to get a better view of both the composition and the geometry of the close environment of R~Scl. We locate the dust emission region between 20 and 90\,mas~($\sim$~3.8--17.0~$R_\star$), as well as the presence of molecular shell around 5--10\,mas~($\sim$~1--2~$R_\star$) and a distinct one around 12--18\,mas~($\sim$~2.3--3.4~$R_\star$). This distinct layer could be made of both: dust and molecules. To reconstruct the formation history of such layer would require a monitoring of the source. Our results are in qualitative agreement with the predictions of dynamic models \citep{Paladini2009} and a detailed comparison with such models is the subject of further study.
   \end{enumerate}

\begin{acknowledgements}

MATISSE has been built by a consortium composed of French (INSU-CNRS in Paris and OCA in Nice), German (MPIA, MPIfR and University of Kiel), Dutch (NOVA and University of Leiden), and Austrian (University of Vienna) institutes. It was defined, funded and built in close collaboration with ESO. The \emph{Conseil Départemental des Alpes-Maritimes} in France, the Konkoly Observatory and Cologne University have also provided resources to manufacture the instrument.
The results are based on public data released from the MATISSE commissioning observations at the VLT Interferometer under Programme IDs 60.A-9257(E).
We have great remembrance of the contributions from our two deceased colleagues, Olivier Chesneau and Michel Dugué to the MATISSE instrument.
We acknowledge with thanks the variable star observations from the AAVSO International Database contributed by observers worldwide and used in this research.
This work was supported by the \emph{Action Spécifique Haute Résolution Angulaire} (ASHRA) of CNRS/INSU co-funded by CNES, as well as from the \emph{Programme National de Physique Stellaire} (PNPS) from CNRS. It has also been supported by the French government through the UCA-JEDI Investments in the Future project managed by the National research Agency (ANR) with the reference number ANR-15-IDEX-01 and the Hungarian governent through NKFIH grant K132406. Vincent Hocdé is supported by the National Science Center, Poland, Sonata BIS project 2018/30/E/ST9/00598. FM acknowledges support from the French National Research Agency (ANR) funded project PEPPER (ANR-20-CE31-0002).

The authors made extensive use of the Jean-Marie Mariotti Center (JMMC) tools to prepare this paper. We can cite among others ASPRO\footnote{\href{http://www.jmmc.fr/english/tools/proposal-preparation/aspro/}{Link to ASPRO on www.jmmc.fr}} used to prepare observations, and OIdb\footnote{\href{http://oidb.jmmc.fr/collection.html?id=12803c68-6124-40ad-803c-68612470ad3d}{Link to OIdb: oidb.jmmc.fr}} to publish the reduced data.

The authors are very grateful to the anonymous referee for his/her careful review of the manuscript, resulting in insightful recommendations and constructive suggestions leading to the great improvement of its content and presentation. We are also very grateful to Orlagh Creevey for her precious information regarding the Gaia distances of R~Scl, and Bernhard Aringer for the fruitful discussions about the \texttt{COMARCS} models. 

\end{acknowledgements}

\bibliographystyle{aa}
\bibliography{bibRScl}

\begin{appendix}

\onecolumn

\section{Observational Information}

\subsection{Log of the MATISSE observations} \label{sec:logobs}

\begin{table}[htbp]
\centering
\begin{minipage}[htbp]{11cm}
\caption[]{Log of the MATISSE observations of R~Scl.}
\label{tab:log}
\centering
 {\renewcommand{\arraystretch}{1.2}
\begin{tabular}{l|l|c|c|c|c}
   \renewcommand{\footnoterule}{} 
Observing date & VLTI config.  & $BLR$  & $N/N_{\star}$ & Seeing & $\tau_0$ \\
\hline
\hline
2018-12-03 $\rightarrow$ 05  & A0--B2--C1--J2     & 10--129     &  9/14 & 0.94 & 0.6 \\
2018-12-07 $\rightarrow$ 08   & A0--B2--C1--D0 & 9--34      &  6/7 & 0.94 & 2.3\\
2018-12-09   & B2--D0--J3--K0     & 29--137     &  5/5 & 0.73 & 4.1 \\
2018-12-10 $\rightarrow$ 11    & D0--G2--J3--K0 & 34--104     &  10/10  & 0.82 & 4.0\\
2018-12-12   & A0--D0--G1--J3     & 25--131     &  4/4 & 0.80 & 4.6\\
2018-12-13 $\rightarrow$ 14  & A0--G1--J2--J3& 39--132 &  9/9 & 0.67 & 5.0\\
2018-12-15   & A0--G1--J2--K0
& 48--123     &  2/2 & 0.53 & 4.7 \\
\hline
\hline
\end{tabular}
}
\begin{tablenotes}
  \small
  \item Notes: $BLR$ stands for Baseline Length Range expressed in~m covered with each VLTI configuration. The ratio $N/N_{\star}$ is the ratio of the used files to the pointings to the star, the seeing in arcseconds, and $\tau_0$ is the coherence time in~ms.
\end{tablenotes}
\end{minipage}
\end{table}

\subsection{List of calibrators} \label{calib}

\begin{table*}[htbp]
\centering
\begin{minipage}[t]{12cm}
\caption[]{List of the calibrator targets used for R~Scl sorted by their decreasing effective temperature.}
\label{tab:calib}
\centering
 {\renewcommand{\arraystretch}{1.2}
\begin{tabular}{l|c|c|c|c|c}
   \renewcommand{\footnoterule}{} 
Name  & Sp. Type & $T_{\rm eff}$~(K)  & $\theta_{\rm UD}$~(mas) & $F_{\rm L}$~(Jy)  & $F_{\rm N}$~(Jy)  \\
\hline
\hline
\object{HD 10144}  & B6V   & $15\,216\pm100$\protect\footnote{\citet{Chandler2016}}    &  $1.8\pm0.2$ & $120\pm14$ & $18\pm4$ \\
\object{HD 48915} & A1V    & $9\,711\pm23$\protect\footnote{\citet{Huang2015}}     &  $6.2\pm0.5$ & $996\pm55$ & $118\pm14$ \\
\object{HD 6805} & K2-IIIb  & $4\,687\pm85$\protect\footnote{\citet{Sousa2018}}     &  $3.3\pm0.3$ & $135\pm3$ & $20\pm2$ \\
\object{HD 39425}      & K1III & $4\,631\pm120$\protect\footnote{\citet{Stevens2017}}     &  $3.7\pm0.4$ & $178\pm24$ & $23\pm4$ \\
\object{HD 32887}   & K4III & $4\,243\pm25$\protect\footnote{\citet{Jofre2015}} &  $6.0\pm0.6$ & $403\pm42$ & $56\pm7$ \\
\object{HD 23614}  & M1III &  $3\,841\pm100^{a}$  &  $5.3\pm0.5$ & $199\pm36$ & $28\pm4$\\
\object{HD 1038}       & M1III & $3\,802\pm100^{a}$     &  $5.6\pm0.6$ & $255\pm45$ & $41\pm10$\\
\object{HD 51799} & M1III &  $3\,780\pm100^{a}$      &  $4.7\pm0.6$ & $183\pm38$ & $29\pm6$\\
\object{HD 11695}      & M4III &  $3\,550\pm50^{b}$     &  $9.3\pm1.2$ & $224\pm198$ & $69\pm19$\\
\hline
\hline
\end{tabular}
}
\begin{tablenotes}
  \small
  \item Notes: The values of the uniform-disk angular diameter are reported in the II/346 VizieR catalog \citep{Bourg2014}, those of the $L$- and $N$-band fluxes in the II/361 catalog \citep{cruzalebes2019}.
\end{tablenotes}
\end{minipage}
\end{table*}

\newpage

\subsection{(u,v)-plane coverage} \label{uvplane}

\begin{figure*}[htbp]
\begin{center}
\includegraphics[width=0.5\textwidth]{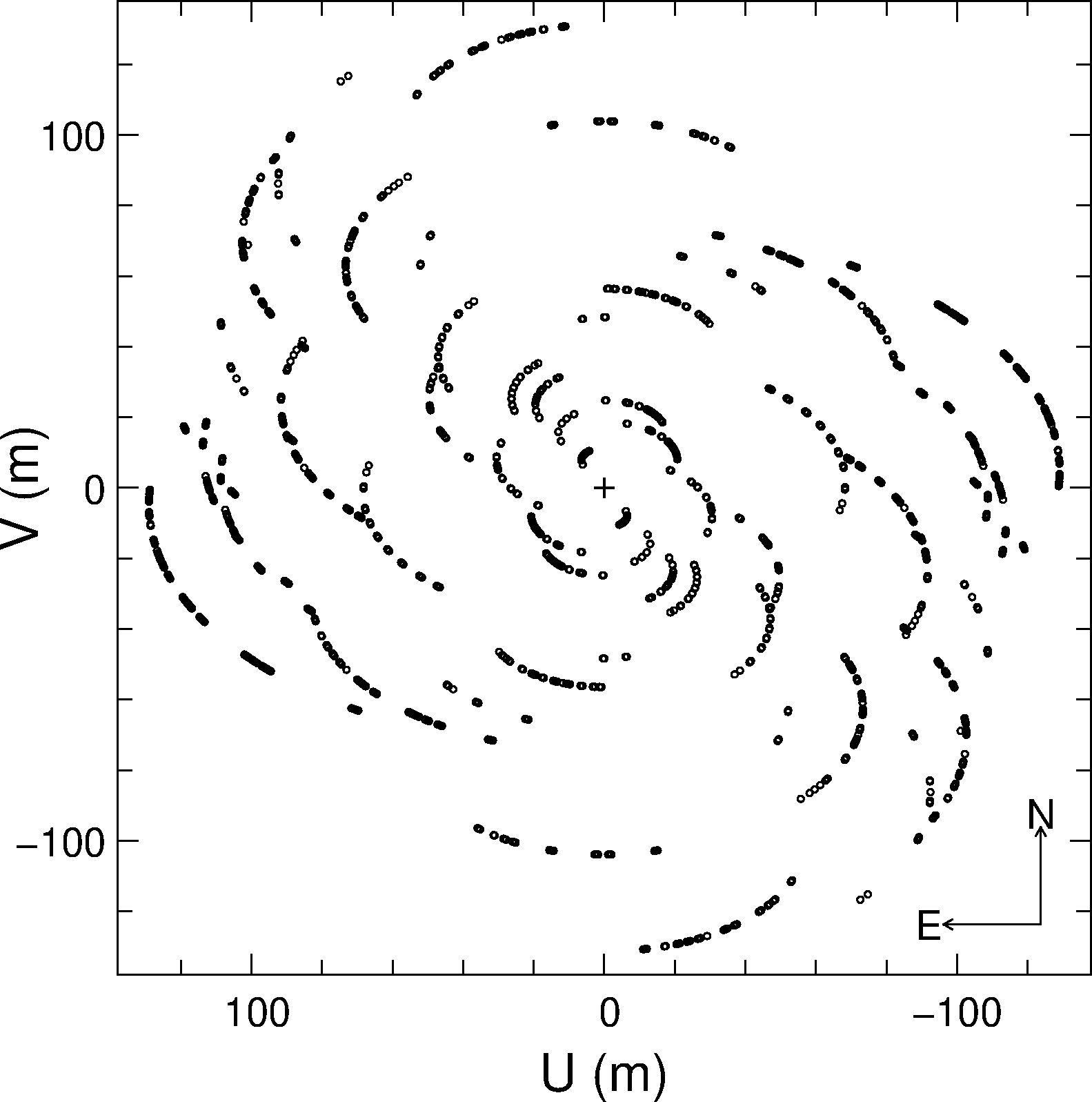}
\caption{Final (u,v)-plane coverage on R~Scl with VLTI/MATISSE. The U and V coordinates are those of the projected baseline vectors.}
\label{fig:uvcov}
\end{center}
\end{figure*}

\subsection{Photometric data} \label{photom}

\begin{table*}[htbp]
\begin{minipage}[t]{8cm}
    \caption{List of the photometric data used in this work (see text for details).}
    \label{tab:photometric}
    \centering
    \begin{tabular}{c | c | c | c | c | c } 
    Band & MJD & $\varphi$ & $A_{\lambda}$ & Magnitude & Flux (Jy) \\ 
    \hline \hline
    V  & 38~700 & 0.85 & 0.18 & $5.84\pm0.02$ & $20.8 \pm 1.0$ \\
    V  & -- & -- &  --  & $5.76\pm0.02$ & $22.6  \pm 1.1$ \\
    V  & -- & -- &  --  & $5.71\pm 0.02$ & $23.5 \pm 1.2$ \\
    R  & -- & -- & 0.13 & $3.76\pm 0.02$ & $105  \pm 4.2$ \\
    R  & -- & -- &  --  & $3.72\pm 0.02$ & $108  \pm 4.3$ \\
    R  & -- & -- &  --  & $3.68\pm 0.02$ & $112  \pm 4.5$ \\
    I  & -- & -- & 0.07 & $2.38\pm 0.03$ & $345  \pm 10.4$ \\
    I  & -- & -- &  --  & $2.35\pm 0.03$ & $318  \pm 9.6$ \\
    I  & -- & -- &  --  & $2.29\pm 0.03$ & $323  \pm 9.7$ \\
    \hline
    J  & 52~888 & 0.98 & 0.05 & $1.64 \pm 0.03$ & $371 \pm 11$ \\
    H  &   --  &  --  & 0.03 &  $0.39 \pm 0.03$ & $779 \pm 23$ \\
    K  &   --  &  --  & 0.02 & $-0.23 \pm 0.03$ & $843 \pm 25$ \\
    L  &   --  &  --  & 0.01 & $-0.81 \pm 0.05$ & $614 \pm 18$ \\
    \hline
    J  & 50~274 & 0.95 & 0.05 &  $1.69 \pm0.03$ & $354 \pm 10$ \\
    H  &   --  &  --  & 0.03 &  $0.45 \pm 0.03$ & $736 \pm 22$ \\
    K  &   --  &  --  & 0.02 & $-0.14 \pm 0.03$ & $777 \pm 23$ \\
    L  &   --  &  --  & 0.01 & $-0.66 \pm 0.05$ & $535 \pm 16$ \\
\hline \hline
\end{tabular}
\begin{tablenotes}
  \small
  \item Notes: $\varphi$ is the pulsation phase and $A_{\lambda}$ is the extinction value given with relative errors of 25\%, computed thanks to the tabulated extinction values E(B-V) from \cite{extinction_computation}. Magnitude is the observed magnitude provided by the catalog and the last column Flux is the dereddened flux computed using the $A_{\lambda}$ values.   
\end{tablenotes}
\end{minipage}
\centering
\end{table*}

\newpage

\subsection{Closure phase} \label{closurephase}

\begin{figure*}[htbp]
\begin{center}
\includegraphics[width=0.5\textwidth]{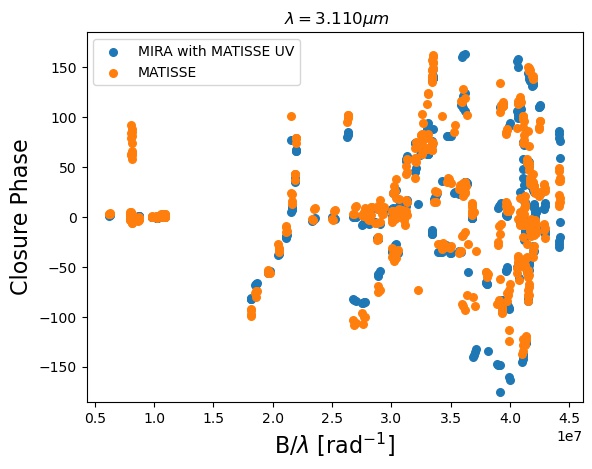}
\caption{Comparison between the closure phase from the MIRA image reconstruction using the MATISSE (u,v)-plane coverage (blue dots) and the closure phase from MATISSE (orange dots).}
\label{fig:closurephase}
\end{center}
\end{figure*}

\subsection{Images convolved by the interferometer beam} \label{sec:convolved}

\begin{figure*}[htbp]
\begin{center}
\includegraphics[width=0.98\textwidth]{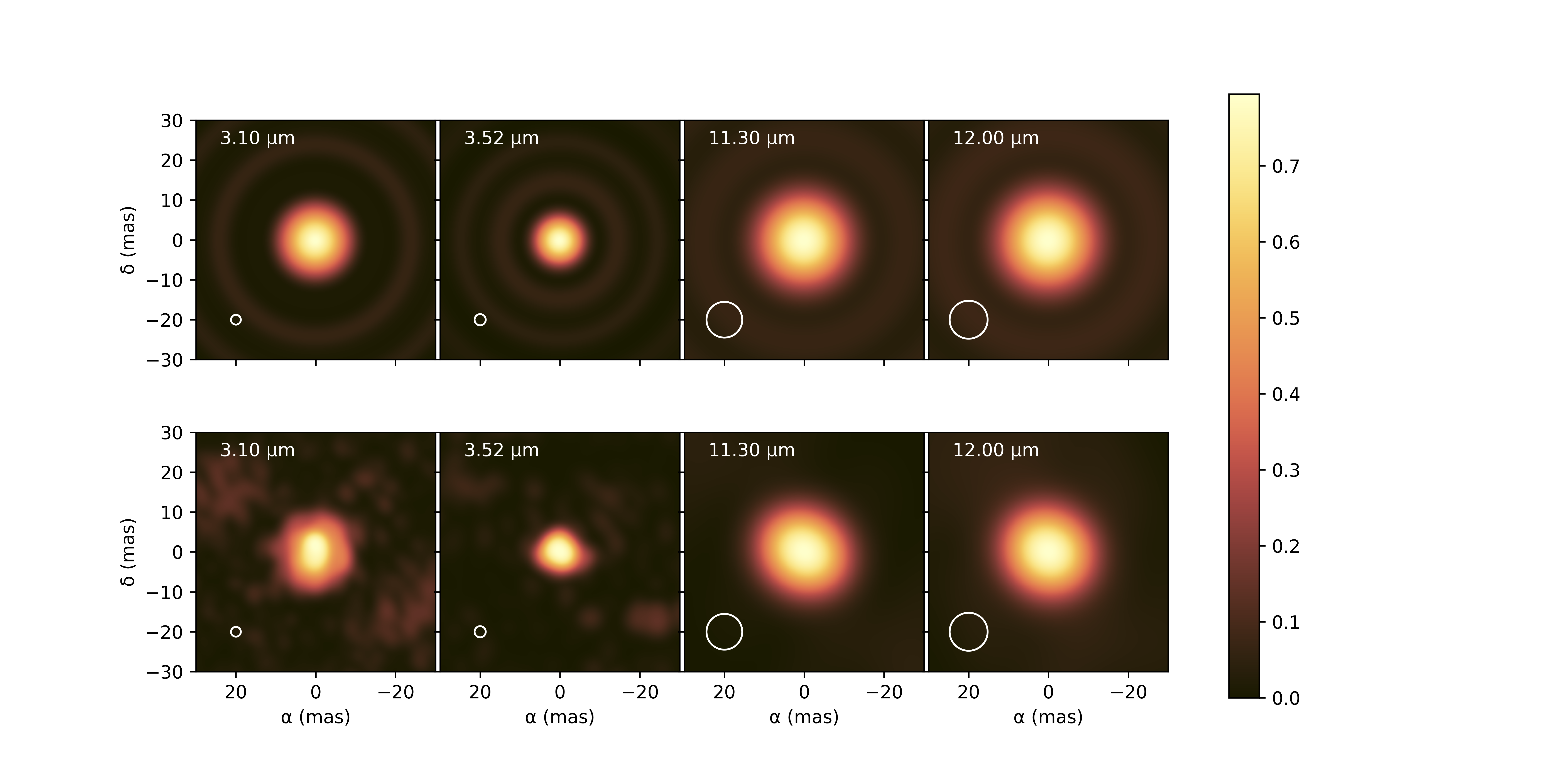}
\caption{Reconstructed Hankel distribution from radial profiles (upper panels) and MIRA images (lower panels) convolved by a Gaussian beam of FWHM the angular resolution of the interferometer.}
\label{fig:images_convol}
\end{center}
\end{figure*}

\newpage 

\subsection{Contour plots} \label{sec:contours}

\begin{figure*}[htbp]
\begin{center}
\includegraphics[width=0.98\textwidth]{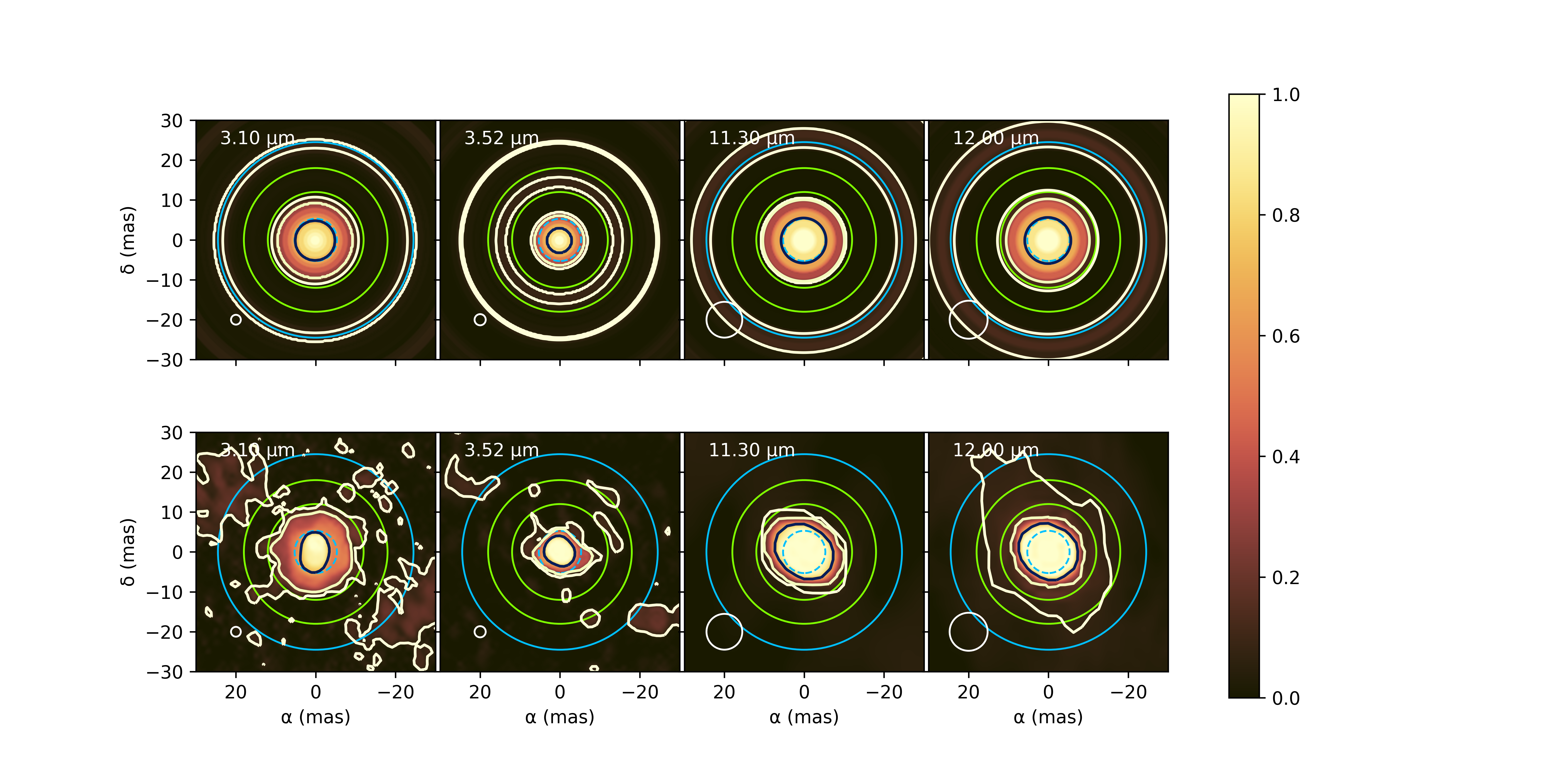}
\caption{Reconstructed Hankel distribution from radial profiles (upper panels) and MIRA images (lower panels) with contours plot in white at 50\%, 5\%, and 0.5\% of the peak intensity level of the images.}
\label{fig:images_contour}
\end{center}
\end{figure*}

\twocolumn

\section{Hankel Profile Reconstructor (\texttt{RHAPSODY})}\label{sec:Hankelprofile}

Here, we describe the method used to reconstruct Hankel profiles in the current work with our Hankel Profile Reconstructor, named \texttt{RHAPSODY,} which stands for:\ Reconstructing Hankel rAdial Profiles in centro-Symmetric Objects with Discrete rings for astrophYsics.

Instead of forcing the routine to describe a given component (e.g., the dust envelope) with a particular geometrical model (e.g., a Gaussian disk), we use a large number of concentric uniform narrow rings (of width given by the angular resolution of the instrument) to reproduce the radial profile of the stellar environment. With this method, thick rings are reproduced by contiguous narrow rings.
This Hankel profile is built thanks to a set of $N_{\rm step}$ concentric circular rings with diameters $\theta_k = k \theta_0$, where $\theta_0 = 0.5 \lambda / B$ is selected as a fraction of the angular resolution of the interferometer. Within that assumption, we can write the normalized monochromatic radial profile as:
\begin{equation}
    \begin{aligned}
    \end{aligned}
    I_{\rm Hankel}( \lambda ) = \frac{4}{\pi} \sum_{k=0}^{N_{\rm step}-1} h_{k}( \lambda )~\frac{ \Pi ( \theta_{k+1} ) - \Pi ( \theta_{k} ) }{\theta_{k+1}^2- \theta_{k}^2},
\end{equation}
where $\Pi (\theta)$ is the rectangular function of width $\theta$.
For physical reasons, the height in intensity $h_{k}$ of each step fulfils the condition $0 \le h_{k} \le h_{0}$ whatever the wavelength.

Thus, the final expression of the global visibility is:
\begin{equation}
    \begin{aligned}
    \end{aligned}
    V_{\rm Hankel} ( q, \lambda ) = \sum_{k=0}^{N_{\rm step}-1} a_{k} ( \lambda )~\frac{ \theta_{k+1}^2~V_{\rm UD} ( q; \theta_{k+1} ) - \theta_{k}^2~V_{\rm UD} ( q; \theta_{k} ) }{\theta_{k+1}^2-\theta_{k}^2},
\end{equation}
where the visibility coefficients $a_{k}$ estimated by our reconstruction method are:
\begin{equation}
    \begin{aligned}
    \end{aligned}
    a_{k} ( \lambda ) = \frac{h_{k} ( \lambda ) }{\sum_{k=0}^{N_{\rm step}-1} h_{k} ( \lambda ) }.
\end{equation}

Adjusting $V_{\rm Hankel}$ to the measured visibilities $V_{\rm obs}$ (with their uncertainties $\sigma_{\rm obs}$) is then a matter of minimizing the distance between both through the $\chi^2$ minimization, varying the $a_{k}$ coefficients.

However, trying to adjust the visibility deduced from a lot of regular intensity steps using a bare final $\chi^2$ value usually gives poor results, because the number of parameters $a_k$ to fit outperforms rapidly the number of available data, in a similar way as in an image reconstruction problem. This is where the Bayesian method comes into the play by adding to the simple $\chi^2$ an a priori bit of information. This additional information to the problem-solving is called a regularization term, $f_{\rm prior}$, calculated on the variable parameters themselves. Such a regularization term must be designed so as to maximize the probability that the reconstructed profile resembles the actual one. For more details on how to deal with heavy inverse problems in a Bayesian  framework, we refer to the excellent introduction to image reconstruction for optical interferometry \citet{Thiebaut2010}.

Adjusting the Hankel profile to the visibility data is then a matter of adding the $\chi^2$ and the regularization term $f_{\rm prior}$ multiplied with a so-called hyperparameter\emph{} $\mu$, which is arbitrarily set:

\begin{equation}
f_{\rm tot} ( \lambda ) = \chi^2 ( \lambda ) + \mu~f_{\rm prior} ( \lambda )
\label{eq:bayes}
.\end{equation}

In our case, we tested two different regularization terms for $f_{\rm prior}$: (1) a smoothness term, $f_{\rm prior} = f_{\rm smooth}$, or (2) a total variation term, $f_{\rm prior} = f_{\rm var}$. On the one hand, $f_{\rm smooth}$, described by the intensity difference between adjacent rings, is given by:
\begin{equation}
    \begin{aligned}
    & f_{\rm smooth} ( \lambda ) = \sum_{k=0}^{N_{\rm step}-1} \sqrt{\bigg[ h_{k+1} (  \lambda ) - h_k ( \lambda ) \bigg]^2}. \\
    \end{aligned}
\end{equation}

We note that $f_{\rm smooth}$ will tend to minimize the variations between consecutive radii, resulting into a smooth radial profile. On the other hand, $f_{\rm var}$ is defined as:
\begin{equation}
    \begin{aligned}
    & f_{\rm var}( \lambda ) = \sum_{k=0}^{N_{\rm step}-1} \left[ \frac{h_{k+1} (  \lambda ) - h_k ( \lambda )}{ \theta_{k+1}-\theta_{k}} \right]^2, \\
    \end{aligned}
\end{equation}
and will tend to reconstruct uniform intensities with sharp edges. Solving the problem (i.e., finding the right radial profile) consists therefore of minimizing the term in Eq.~\ref{eq:bayes}, by varying the parameters $a_k$ using standard \texttt{Cobyla} minimization routines from the \texttt{Lmfit} \footnote{made freely available at \url{https://github.com/lmfit/lmfit-py/}} library, found in \texttt{Python}.

\section{RHAPSODY dynamic range and detection limit} \label{sec:rhapsody_detection}

In this section, we describe how we tested the robustness of the \texttt{RHAPSODY} code by determining its intrinsic dynamic range. To achieve such estimation, we built a toy model mimicking R~Scl environment. Such model is composed of: \texttt{DUSTY} and a gaussian layer at 15\,mas with an intensity ratio with respect to the central source starting from 1/1000 and reaching up to 1/10 (Fig~\ref{fig:rhapsody_test}).
Using \texttt{ASPRO2}, we generated VLTI/MATISSE visibilities, added error noise to the data, and used the same (u,v)-plane coverage as the one of our R~Scl observations (see Fig~\ref{fig:uvcov}) for each of our different models and for both $LM$- and $N$-bands. 
The initial guess intensity profiles used by \texttt{RHAPSODY} for all the simulations in both bands are similar to the one used in this paper (i.e., Gaussian profile with a $FWHM= 7$\,mas). The value of the hyperparameter used is $\mu=1$. 
Figure.~\ref{fig:rhapsody_test} shows the spectra of the intensity profile reconstruction produced by \texttt{RHAPSODY}. We can notice that the lower limit of the dynamic range in $L$-band is about 1/200 (0.5\%). In $N$-band, the code does not reconstruct (or does so very marginally) the inner Gaussian structure added at 15\,mas, whatever its intensity. We suspect this non-detection in the $N$-band is due to an unfortunate combination of instrumental angular resolution and dynamical range.

\begin{figure*}[htb]
    \centering

    \subfloat[Without the gaussian layer at 15\,mas]{    
    \includegraphics[width=0.9\textwidth]{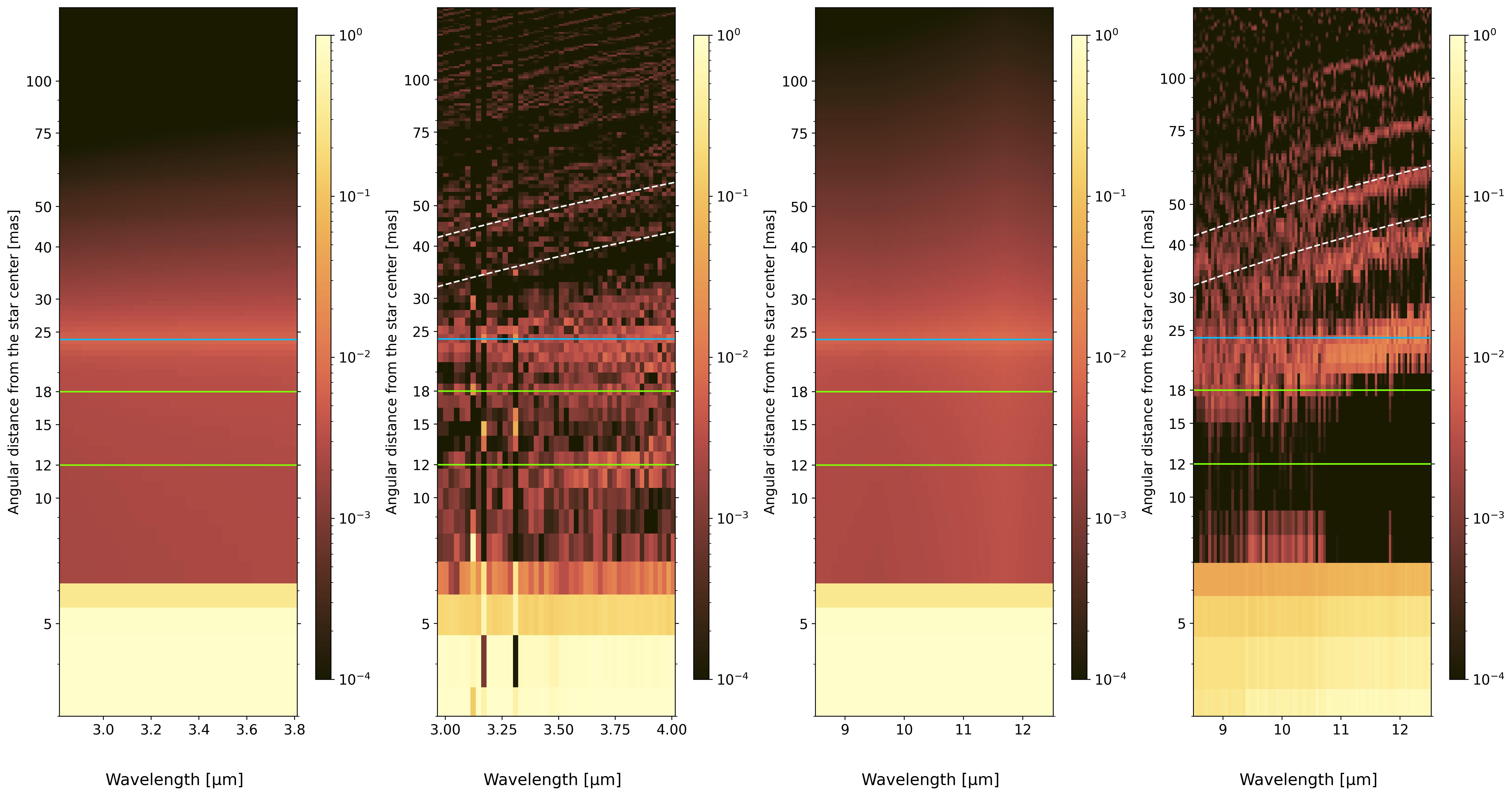}
    }
    \hfill

    \subfloat[Additional gaussian layer at 15\,mas with an intensity ratio with respect to the star of 1/1000]{    
    \includegraphics[width=0.9\textwidth]{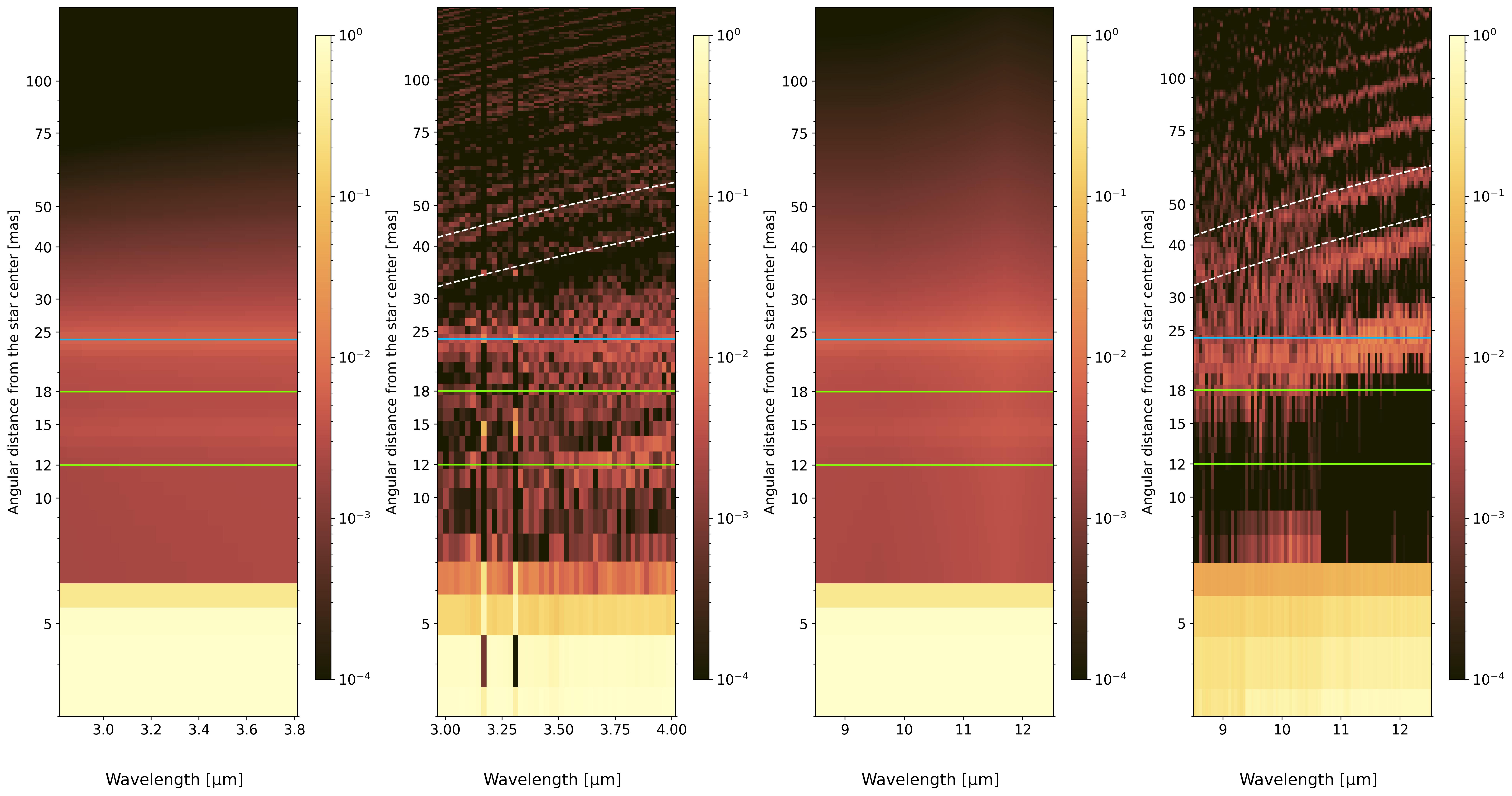}
    }
    \hfill

    \caption{continued.}
    \label{fig:7}
\end{figure*}

\begin{figure*}[htb]
    \ContinuedFloat
    \centering

    \subfloat[Additional gaussian layer at 15\,mas with an intensity ratio with respect to the star of 1/200]{    
    \includegraphics[width=0.9\textwidth]{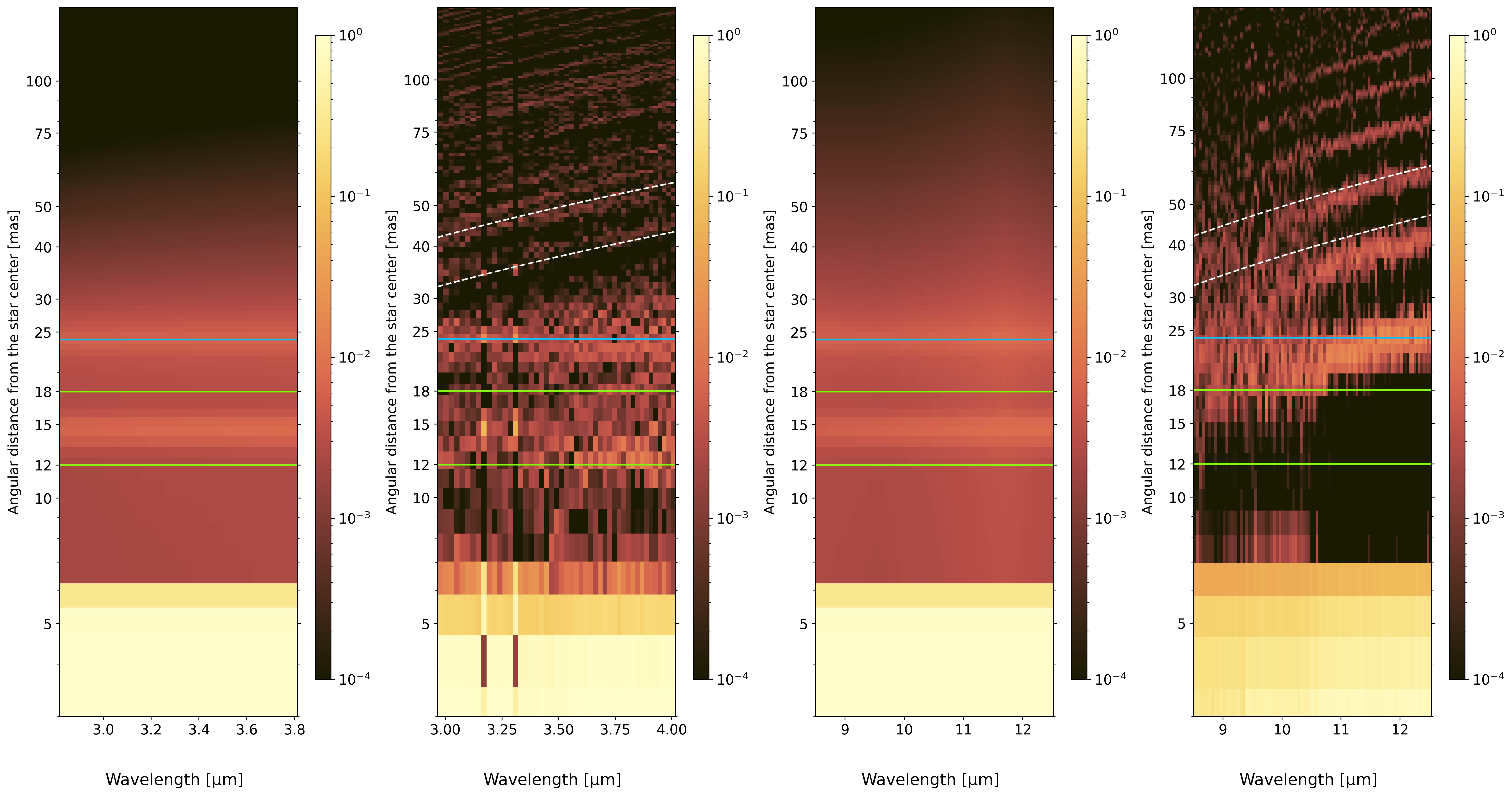}
    }
    \hfill

    \subfloat[Additional gaussian layer at 15\,mas with an intensity ratio with respect to the star of 1/100]{    
    \includegraphics[width=0.9\textwidth]{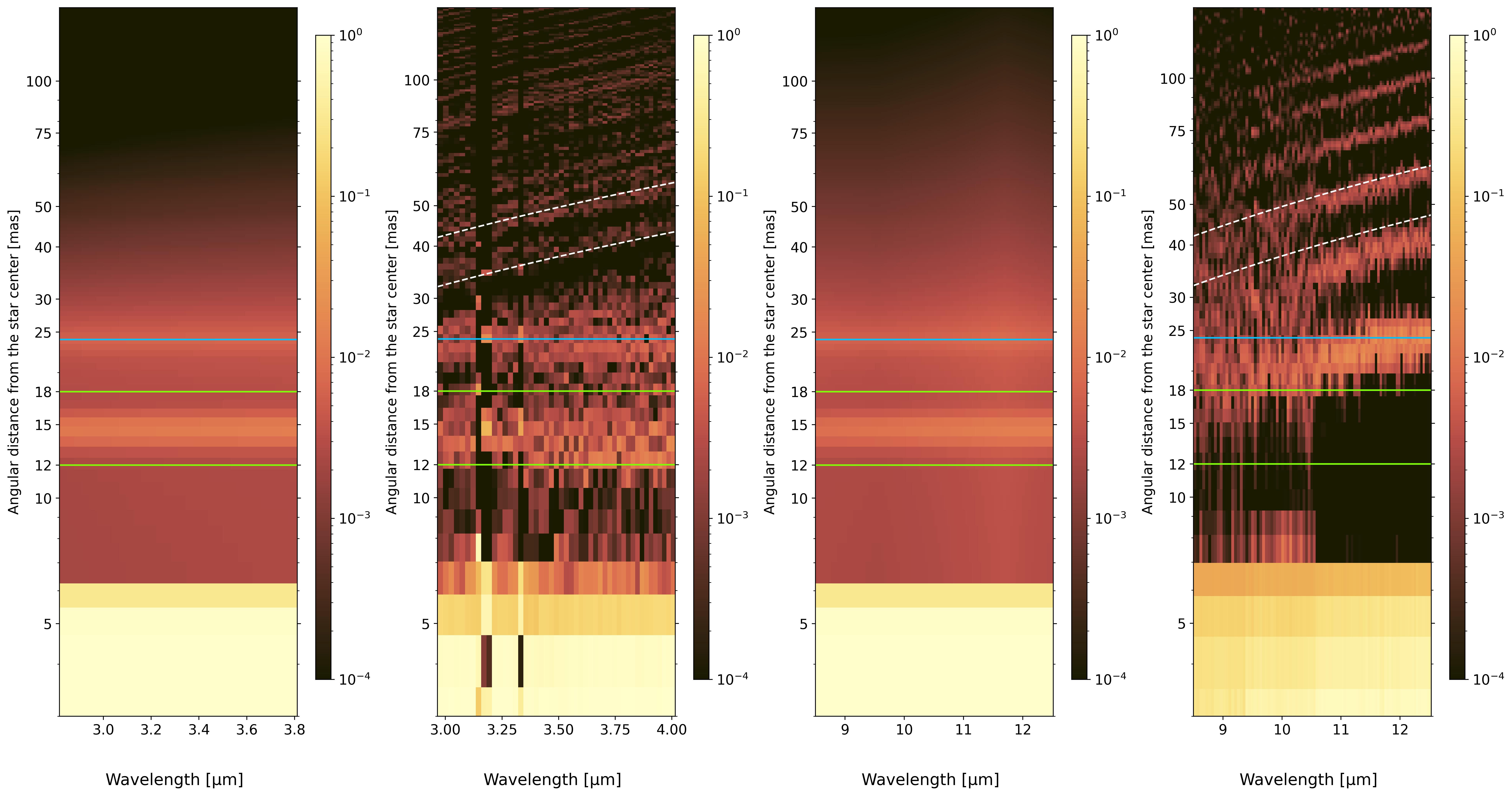}
    }
    \hfill

    \caption{continued.}
    \label{fig:7b}
\end{figure*}

\begin{figure*}[htb]
    \ContinuedFloat
    \centering

    \subfloat[Additional gaussian layer at 15\,mas with an intensity ratio with respect to the star of 1/10]{    
    \includegraphics[width=0.9\textwidth]{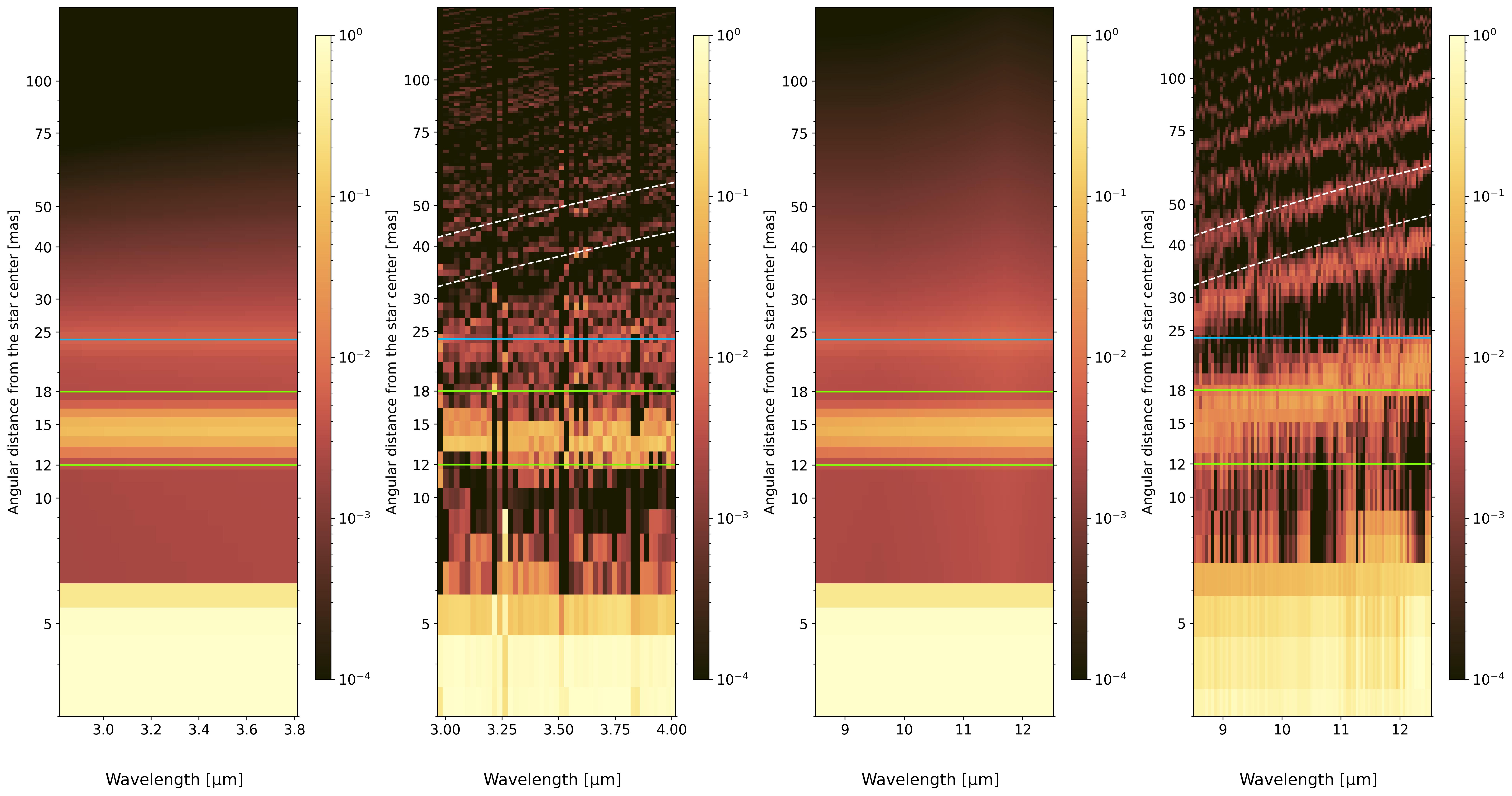}
    }
    \hfill

    \caption{Modeled intensity spectra from \texttt{DUSTY} panels 1 and 3 and its associated \texttt{RHAPSODY} intensity profiles reconstruction spectra 2 and 4 in the $L$-Band and $N$-band, respectively (from left to right). A gaussian layer has been also added at 15\,mas with an intensity ratio compared to the star of 1/1000, 1/200, 1/100, 1/50, 1/20, and 1/10.}
    \label{fig:rhapsody_test}
\end{figure*}

\end{appendix}

\end{document}